**Early warning of road icing from antecedent meteorological conditions without consistent critical slowing down**

**Chaeyeon Yi**[1,3] · **Yun Am Seo**[2,3,*]

[1] Research Center for Atmospheric Environment, Hankuk University of Foreign Studies, Yongin 17035, Republic of Korea

[2] Department of Data Science, Jeju National University, Jeju 63243, Republic of Korea

[3] NAVI Hyper-Tropicalization Research Institute, Jeju National University, Jeju 63243, Republic of Korea

[*] Corresponding author · seoya@jejunu.ac.kr

ORCID Chaeyeon Yi 0000-0003-2802-2431 · Yun Am Seo 0000-0001-9283-4376

**Abstract**

Critical slowing down (CSD), the slowing of recovery from perturbations as a system approaches a critical transition, is widely used for early warning, but whether such signals appear in environmental transitions subject to rapidly changing external forcing has rarely been tested with observations. Here we show, using 2,427 icing-onset events in 1-min observations from 121 South Korean road weather stations, that such signals do not consistently precede road icing. Icing began from different prior surface states, but conditions at onset varied continuously with weak separation among pathways. Cold dose, deposition dose and freeze–thaw accumulation at onset exceeded those at non-icing control times, with cold-dose discrimination highest at decay timescales of 30 min or less. However, the increases in variance and lag-1 autocorrelation expected under CSD did not distinguish icing from control times; the same procedure recovered these increases in synthetic series with a declining recovery rate, and the null result was replicated for 3,091 events at 127 independent stations. Predictive information was nevertheless present: adding recent changes and cumulative antecedent conditions to current conditions improved discrimination, and surrounding meteorological fields increased the area under the ROC curve by 0.030–0.032 for forecast windows of 1 to 12 h and reduced false alarms by 62–67% at 3 and 6 h. These results suggest that road icing is more consistent with recent meteorological forcing moving the road surface into icing-favourable conditions than with a progressive decline in internal stability, and they indicate where early warning should look when a transition is externally forced.

## Introduction

Many complex systems, including ecosystems and the climate system, can undergo critical transitions, in which the system state shifts abruptly while external forcing changes gradually [1]. As a system approaches a bifurcation at which it loses stability, recovery to the original state after a perturbation can slow down, a phenomenon known as critical slowing down (CSD). This loss of stability can appear in time series as increases in variance and lag-1 autocorrelation, which have been used as standard early warning signals (EWS) of critical transitions [2–4]. CSD-based early warning research has expanded beyond ecosystems and the climate system to fields such as financial markets [5] and traffic flow [6]. To our knowledge, however, CSD-based early warning has not been tested for transitions in road-surface state, including the onset of icing.

Not all abrupt state transitions arise from the approach to a bifurcation. Transitions can result from different dynamical processes, including a gradual loss of internal stability, stochastic perturbations and the rate of change of external forcing [7]. When changing external forcing moves a system across a threshold without a progressive loss of stability near a bifurcation, CSD is not necessarily expected to precede the transition. State transitions without clear CSD have indeed been reported [8], and uncertainties in detecting and interpreting CSD indicators in observational data, arising from record length, noise and the detrending method, have also been noted [9,10]. Nonstationary noise can mask critical slowing down or produce spurious warning signals, and alternative indicators proposed to be robust to it have been evaluated in idealized models of tipping elements [11].

This distinction matters for early warning of real environmental hazards. For a phenomenon that approaches a bifurcation through a gradual decline in internal stability, tracking internal precursors such as the variance or autocorrelation of the system itself may be effective. For transitions governed more directly by changing external conditions, however, the forcing itself and its temporal accumulation may contain more useful predictive information than internal precursors. Testing CSD together with antecedent forcing in real observations can therefore help distinguish where predictive information resides before a transition.

Road icing is a major winter road-weather hazard because icy surfaces reduce friction and can be difficult for drivers to detect, particularly when ice is thin and transparent [12–16]. Despite these hazards, large uncertainty remains about when a non-icing road surface actually transitions to an icing state.

A drop in road surface temperature to or below the freezing point is an important condition that raises the likelihood of icing, but it alone cannot explain the actual occurrence of icing. In 48,060,381 valid observation minutes, road surface temperature was at or below 0 °C for 27.2% of the time, but icing occurred in only 1.5% of those subfreezing observations. A previous study of the same observation network also found that the likelihood of icing increased gradually under combined low-temperature and high-humidity conditions rather than discontinuously at a specific temperature [17]. Road surfaces also have thermal characteristics distinct from those of natural surfaces because they are affected by heat storage in the pavement and subsurface, the radiation budget and heat exchange with the atmosphere [18–20]. A recent study based on a road weather information system (RWIS) showed that a road surface temperature at or below 0 °C is an important but not sufficient condition for icing [21].

RWIS observations provide the conditions required for this test. Fixed RWIS stations observe road surface temperature, meteorological variables and surface state continuously at short intervals, and at some stations ice-film sensor information can be used as an auxiliary check on surface-state observations.

Research on the occurrence and prediction of road icing has developed along three main lines: classification of formation mechanisms, physically based prediction of road-surface state, and statistical and machine-learning prediction.

First, mechanism-oriented studies have distinguished formation pathways such as radiative cooling or frost, refreezing after melting, icing after precipitation and freezing precipitation, according to the moisture supply and cooling processes required for icing. Freezing precipitation, including freezing rain and freezing drizzle, has been studied in comparatively fine detail with a focus on the thermodynamic structure of the atmosphere and near-surface temperature conditions [22,23]. Such classifications are useful for explaining that icing does not arise from a single cause. A previous study using part of the data from the same observation network also defined several formation types from meteorological and road-surface conditions before icing and analysed differences among these types [17]. However, the fact that events can be classified by predefined rules does not imply that observed icing events form discrete pathways that are clearly distinguishable from one another. Whether icing events that begin from different antecedent conditions actually occupy distinct regions of state space, or converge on similar conditions at icing onset through continuous changes in conditions, requires a separate test.

Second, physically based models explicitly represent the surface energy budget, heat transfer, moisture and phase changes and are widely used in operational road-weather forecasting [24–27]. Their performance, however, depends on meteorological forcing, initial road-surface conditions and process representation [28]. Their purpose therefore differs from directly testing whether statistical precursors of declining internal stability are detectable in high-frequency observations before icing onset.

Third, statistical and machine-learning studies have sought to discriminate or predict icing or road-surface state by learning relationships between meteorological and road-surface observations and forecast data [29–31]. In addition, because road-weather observations at the same station and in the same winter are strongly dependent in space and time, splitting the data randomly can overestimate predictive performance for new stations or new periods.

We treated road icing as a state transition from a non-icing to an icing state and tested in real observations whether changes before icing onset show characteristics consistent with critical slowing down or whether recent meteorological forcing provides more direct predictive information about icing onset. To this end, we posed three research questions (RQs).

RQ1. Does the process leading to icing form distinguishable pathways, or does it show continuous variation?

RQ2. Do early warning signals consistent with critical slowing down appear before icing onset?

RQ3. How much predictive information do recent and surrounding meteorological conditions provide about future icing onset?

We addressed these questions with 2,427 icing-onset events from 121 RWIS stations on expressways in South Korea and evaluated the reproducibility of the main results with 3,091 icing-onset events at 127 independent RWIS stations with no stations overlapping the primary observation network (Fig. 1a–c; Table 1).

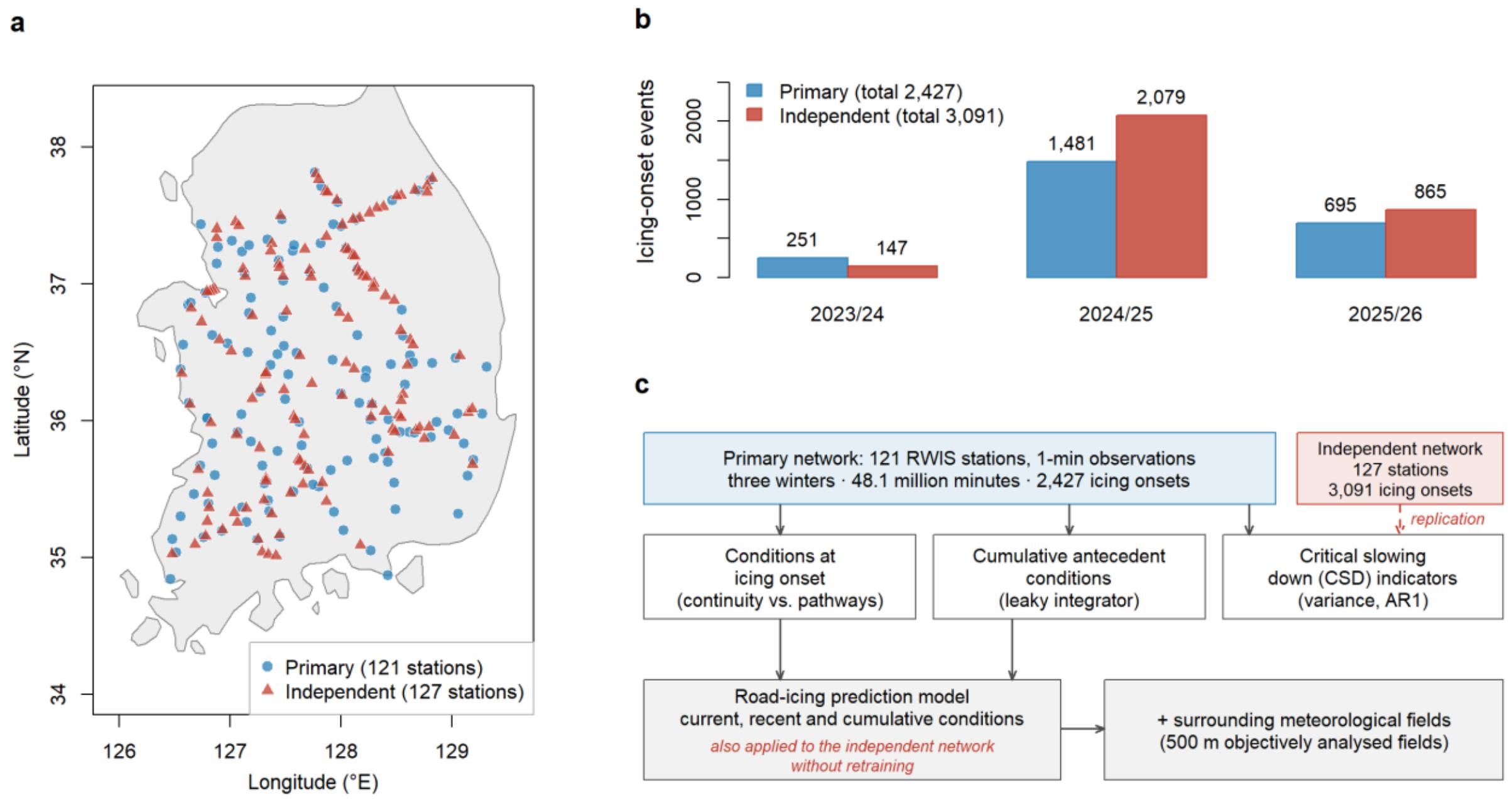

**Fig. 1**

**Table 1**

| Events | Stations | Prior state (WET/SNOW/DRY) | IFT>0 (label agreement) | Winter (2023/24 / 2024/25 / 2025/26) |
|---|---|---|---|---|
| 2,427 | 119 | 1,875 / 380 / 172 | 94.4% | 251 / 1,481 / 695 |

## Results

### Continuous structure and weak pathway separation of road-icing onset conditions

The 2,427 icing-onset events began from different prior surface states: WET was the most common (1,875 events), followed by SNOW (380) and DRY (172). However, the distributions of road surface temperature at onset overlapped widely among prior surface states (Fig. 2a). Prior surface state explained little variation in onset conditions:

$\eta^2$ was 0.044 for road surface temperature, 0.064 for relative humidity and 0.003 for the preceding 24-h temperature range (Fig. 2b).

Separation by prior surface state was also limited in the temporal evolution before icing. The difference in median road surface temperature among events starting from DRY, WET and SNOW increased from 0.7 °C 24 h before icing to 3.2 °C 6–8 h before icing and decreased to 1.9 °C at onset, and road surface temperature was lowest for events starting from SNOW from 24 h before icing until onset (Fig. 2c). When these cooling trajectories were used to classify prior surface state, overall accuracy was 0.809, slightly higher than the 0.782 of a baseline that always selects the most frequent class, and balanced accuracy was 0.486. Antecedent cooling therefore contained some information that distinguished prior surface states, but this information was insufficient to separate the three groups clearly.

The distribution of the first principal component (PC1) showed no clear multimodality (Fig. 2d), and Hartigan's dip test did not reject the null hypothesis of unimodality ($p = 0.99$). Thus, we found no evidence of multimodality along PC1; this test alone does not exclude distinct groups elsewhere in the multivariate feature space. In a supplementary cluster analysis, the mean silhouette coefficient at the optimal $k = 2$ was 0.250, indicating weak separation between clusters, and agreement between the derived clusters and prior surface state was very low (adjusted Rand index, ARI = 0.002).

Meteorological and road-surface characteristics related to icing also changed gradually along PC1 (Fig. 2e). At low PC1, the fraction of the preceding 24 h with observed precipitation, relative humidity at onset and road surface temperature at onset were high and cold dose was low; with increasing PC1, the precipitation fraction ($r = -0.61$), relative humidity ($r = -0.42$) and road surface temperature at onset ($r = -0.79$) decreased, whereas cold dose ($r = 0.73$) increased. In contrast, the 24-h range of road surface temperature ($r = 0.07$) and the time of day of onset ($r = -0.06$) were only weakly related to PC1. Together, these gradients indicate gradual variation in physical conditions along PC1 rather than clearly separated regimes.

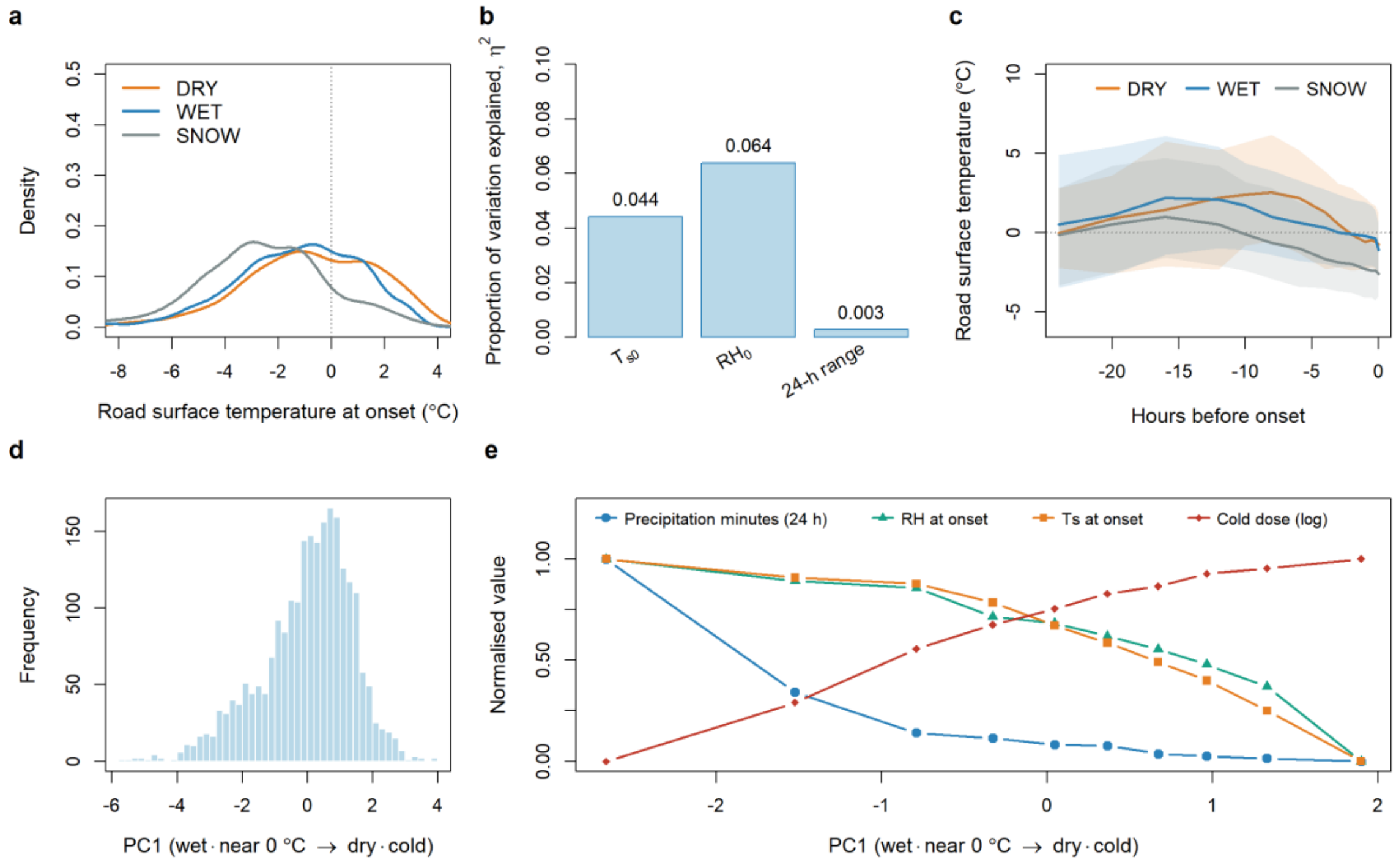

**Fig. 2**

## Accumulation of antecedent conditions and a short decay timescale before road-icing onset

We represented antecedent cooling, conditions favourable for water-vapour deposition and freeze–thaw history using three cumulative indicators: cold dose, deposition dose and freeze–thaw accumulation.

At icing onset, cold dose was distributed towards higher values than at control times (Fig. 3a). Deposition dose also extended to higher values at icing onset (Fig. 3b). Freeze–thaw accumulation was also generally higher at icing onset (Fig. 3c). We also evaluated which decay timescale for past cooling was most effective for distinguishing icing onset (Fig. 3d). When the decay time constant $\tau d$, which sets how quickly the influence of past cooling declines in the cold-dose calculation, was varied from 1 min to 72 h, the area under the receiver operating characteristic curve (AUC) between icing events and non-icing control times was highest at short $\tau d$. At onset, AUC was highest at $\tau d$ = 5 min (0.702), with similar values at 1–10 min. When cold dose was computed 30 and 60 min before onset, excluding onset-time temperature, AUC was highest at $\tau d$ = 20 min (0.675 and 0.671); differences from 10–30 min were 0.002 or less. At all three times, AUC decreased as $\tau_d$ increased beyond 1 h, reaching 0.625–0.628 at $\tau_d$ = 24 h. When discrimination was evaluated within 3-h time-of-day bins to account for the concentration of icing onset at night and in the early morning, AUC remained higher at $\tau_d$ = 5 min (0.675) than at $\tau_d$ = 24 h (0.628). The short $\tau_d$ is an empirical decay timescale at which cold dose discriminated well between icing and non-icing in these observations; it does not represent an intrinsic physical time constant or thermal response time of the road surface.

We also evaluated whether these cumulative antecedent conditions provided predictive information beyond current road-surface and meteorological conditions (Fig. 3e). In station-held-out cross-validation of T+3 h road-icing prediction, AUC was 0.736 with a simple road surface temperature threshold alone and 0.693 with cold dose, deposition dose and freeze–thaw accumulation alone. In contrast, current road surface temperature and relative humidity together gave an AUC of 0.839, which increased to 0.856 when the recent 1-h change in road surface temperature ($\Delta T_s$) was added. Adding the three cumulative antecedent conditions to form the full information set raised AUC to 0.885. Together, recent temperature change and cumulative antecedent conditions increased AUC by 0.047 over the current state, including a further increase of 0.029 from cumulative conditions after $\Delta T_s$ (Fig. 3e). When $\tau_d$ of the cumulative antecedent conditions was changed to 20 or 60 min, AUC of the model using only the cumulative indicators was 0.719–0.729, higher than with $\tau_d$ = 24 h (0.693), whereas AUC of the full model was almost unchanged at 0.883–0.887.

However, this accumulation of antecedent conditions and the associated predictive information do not in themselves imply a decline in the internal stability of the road-surface system. We therefore next tested whether the increases in variance and lag-1 autocorrelation (AR1) expected under CSD appear before icing onset in the observations.

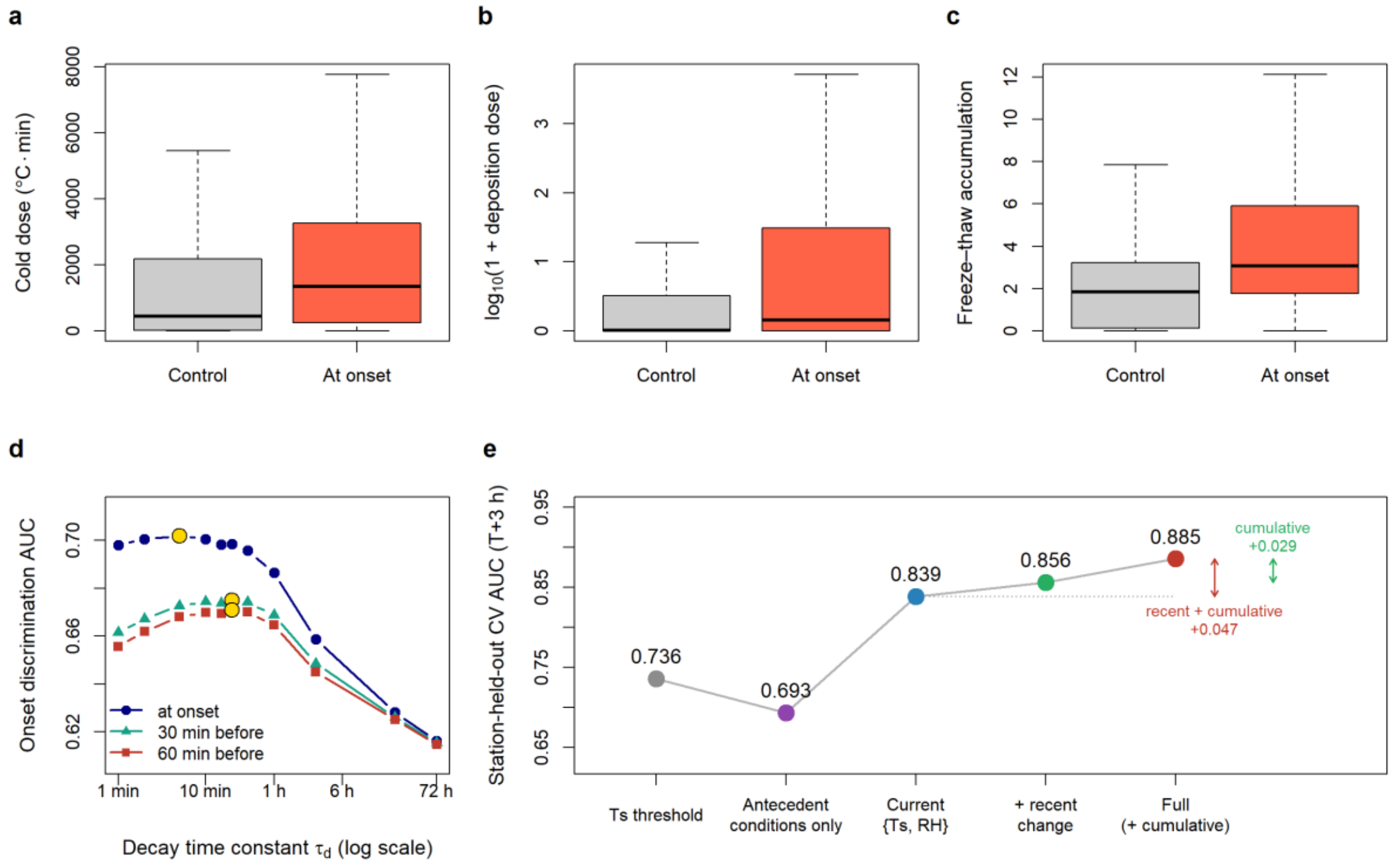

**Fig. 3**

## Absence of CSD-based early warning signals before road-icing onset

In the primary observation network, the 2,427 icing events did not show these increases. Kendall's rank correlation coefficient τK for variance was widely distributed across negative and positive values, and its distributions for icing events and non-icing control times overlapped extensively (Fig. 4a). For AR1, the τK distribution of icing events was not clearly shifted towards positive values and overlapped widely with that of non-icing control times (Fig. 4b). Variance or AR1 increased before some icing events but decreased or showed no clear change before others.

Across nine analysis settings combining analysis-window length and detrending bandwidth, the discrimination AUCs of variance and AR1 were all distributed around 0.5, and no setting showed consistently high discrimination (Fig. 4c). In the primary analysis, AUC was 0.488 for variance and 0.513 for AR1, so the pre-onset trends of these two indicators alone did not distinguish icing events from non-icing control times (Fig. 4d). When six indicators, obtained by adding skewness, kurtosis, spectral reddening and autocorrelation decay time to variance and AR1, were evaluated against random control times and against control times matched for season and time of day, discrimination AUCs also lay around 0.5, ranging from 0.431 to 0.567, and no combination showed consistent discrimination (Supplementary Table 1). Against controls matched for season and time of day the AUCs were slightly higher, 0.548 for variance and 0.560 for AR1, but still close to 0.5.

We used a positive control with synthetic data to check whether the absence of signals in the observations reflected an insufficient detection capability of the analysis procedure. When the same procedure was applied to synthetic series from a linear stochastic process whose recovery rate declined towards the end of the analysis window, the discrimination AUCs of the variance and AR1 trends were 0.618 and 0.636, respectively, for an AR1 increase of 0.045 over 12 h. When the AR1 increase was reduced to 0.020, AUC was 0.549, and for synthetic controls with a fixed recovery rate it was 0.512 (Supplementary Table 2). The same procedure could therefore detect the increase in variance and AR1 generated by a declining recovery rate in the synthetic data used here, whereas comparable increases were not observed in the RWIS data. However, the discrimination AUCs for an AR1 increase of 0.01 (variance 0.508, AR1 0.515) were similar to those of the synthetic controls with a fixed recovery rate (0.512, 0.507), so a weak decline in recovery rate of this magnitude cannot be ruled out with this procedure.

Relaxation-time analysis based on the recovery rate did not show a consistent slowing of recovery before icing either. The relaxation time $\lambda^{-1}$ estimated from AR1 at the end of the analysis window depended strongly on the detrending bandwidth: increasing the bandwidth from 30 to 240 min increased its median from 19 to 113 min. The sign of the difference between icing events and non-icing control times also changed with bandwidth, and a similar sign change appeared in a proxy for the stationary variance. Moreover, the autocorrelation of the detrended residuals did not decay as a single exponential with lag, and relaxation times converted from lag 1 and lag 20 did not agree. The absolute value of $\lambda^{-1}$ therefore could not be reliably interpreted as an intrinsic relaxation time of the road-surface system. Even when trends within the analysis window were evaluated, the discrimination AUCs across the nine analysis settings ranged from 0.43 to 0.56, without consistent discrimination.

Other state variables also lacked the joint variance and AR1 increase expected under CSD (Supplementary Table 3). For deposition margin and air temperature, the AUCs of the variance trend were 0.502 and 0.471, respectively, whereas the AR1 trend gave 0.567 and 0.549, so the two indicators did not increase together. For the friction coefficient, the AUC of the variance trend was 0.683 but that of the AR1 trend was 0.308, in the opposite direction, and for water-film thickness the two indicators also pointed in opposite directions (0.757 and 0.231; Supplementary Table 3). When the trends were recomputed with the window end moved back from onset, the discrimination AUC of friction-coefficient variance decreased from 0.679 with the analysis window ending at icing onset to 0.543 with the window end 12 h earlier (Supplementary Table 4). The variance changes in some surface-state variables therefore differed from the joint increase in variance and AR1 expected under CSD and were concentrated closer to icing onset rather than appearing consistently well in advance.

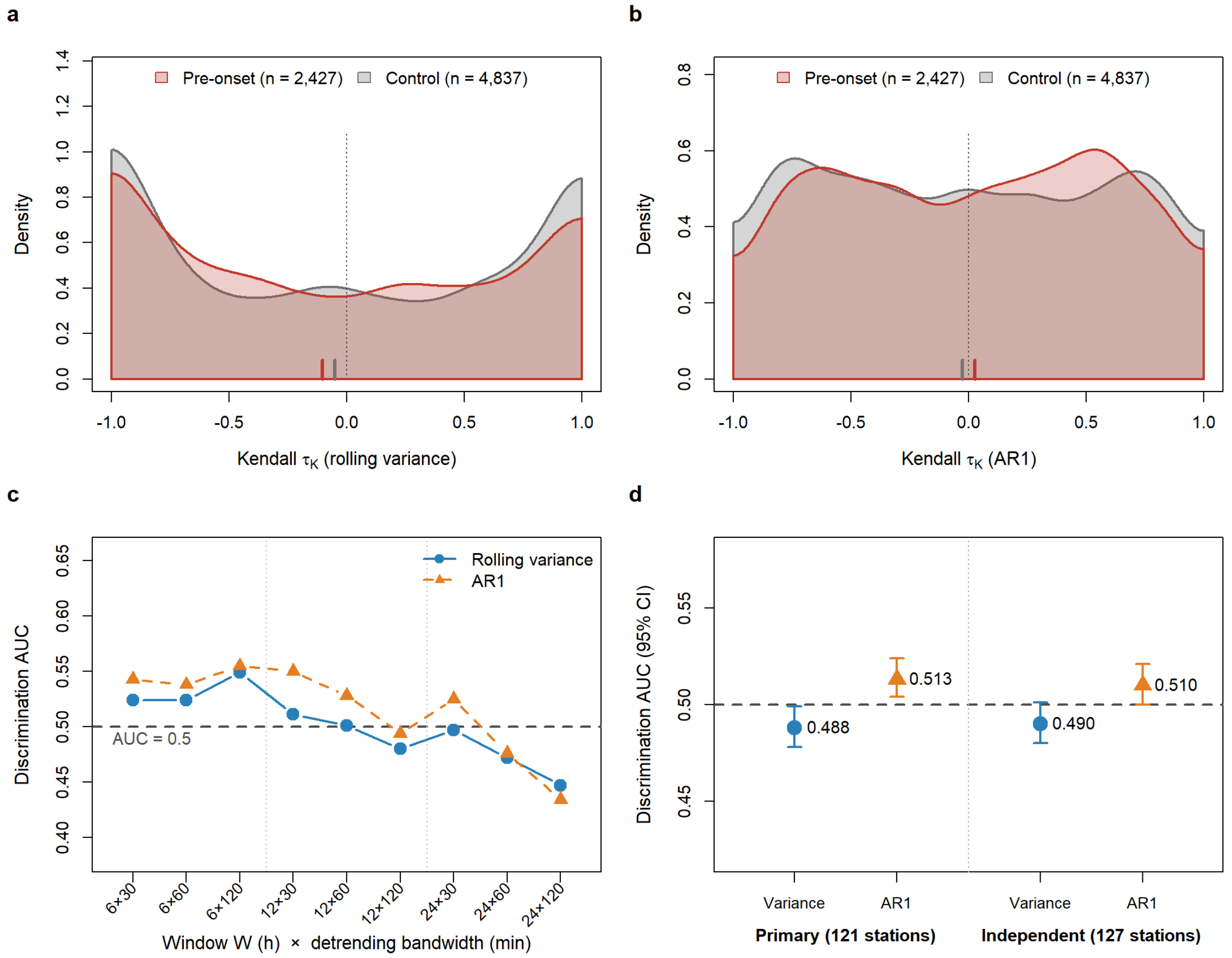

**Fig. 4**

## Independent-network evaluation of CSD signals and icing prediction

To test whether the lack of consistent CSD signals depended on particular stations or data composition, we repeated the analysis for 3,091 icing events at 127 independent

RWIS stations, with no overlap with the 121 primary-network stations (Table 2). The composition of surface states immediately before icing onset was also similar in the two networks, with WET accounting for 77.3% of events in the primary observation network and 76.6% in the independent observation network.

In the independent observation network, variance and AR1 also did not increase consistently on approach to icing onset. The discrimination AUC of the variance trend was 0.488 (95% confidence interval, CI: 0.478–0.499) in the primary observation network and 0.490 (0.480–0.501) in the independent observation network (Table 2; Fig. 4d). The AR1 trend gave similar values in the two networks, 0.513 (0.504–0.524) and 0.510 (0.500–0.521), respectively. When the tolerance for the difference between road surface temperature and air temperature in the independent observation network was tightened from 30 °C to 15 °C and 10 °C, the discrimination AUCs barely changed (variance 0.489–0.490, AR1 0.507–0.510).

Direct tests of the direction of event-level CSD indicator trends were also consistent between the two networks. If CSD were present, variance and AR1 would increase on approach to icing onset, and Kendall's rank correlation coefficient $\tau_K$ would be predominantly positive. However, one-sided Wilcoxon signed-rank tests did not support this direction. For the variance trend, $p$ was > 0.99 in both the primary and independent observation networks ($V$ = 1,303,427, $n$ = 2,427; $V$ = 2,120,289.5, $n$ = 3,089, excluding 2 events whose analysis windows did not meet the data-availability criterion), and for the AR1 trend, $p$ was 0.215 ($V$ = 1,461,318.5) and 0.712 ($V$ = 2,306,715.5), respectively (Table 2). Neither network therefore provided statistical evidence that CSD indicators increase consistently in the expected direction before icing.

These results also did not arise from differences in sample composition caused by the yearly expansion of the networks. When the same analysis was repeated for each winter separately in both networks, the discrimination AUCs ranged from 0.474 to 0.529 for the variance trend and from 0.491 to 0.536 for the AR1 trend, and discrimination was not higher in any particular winter. Including the incompletely observed 2022/23 winter in the analysis period added only two icing-onset events and changed the discrimination AUC by 0.004 or less (Supplementary Table 5), which was smaller than the change produced by redrawing the control times alone (up to 0.009).

Predictive information was nevertheless retained when the road-icing prediction model trained on the primary observation network was applied to the independent observation network without retraining. ROC-AUC was 0.849 for T+3 h and 0.820 for T+6 h prediction; at a probability of detection (POD) of 0.80, false alarms were 6.59 and 6.32 per station-day, respectively (Supplementary Table 6).

**Table 2**

| | **Primary set (121 stations)** | **Independent set (127 stations)** |
|---|---|---|
| Events / stations | 2,427 / 119 | 3,091 / 127 |
| Prior state WET (%) | 77.3% | 76.6% |
| Rolling-variance $\tau_K$ AUC (95% CI) | 0.488 (0.478–0.499) | 0.490 (0.480–0.501) |
| Lag-1 AR $\tau_K$ AUC (95% CI) | 0.513 (0.504–0.524) | 0.510 (0.500–0.521) |
| One-sided Wilcoxon signed-rank test for variance $\tau_K$ > 0 | p > 0.99<br>(V = 1,303,427; n = 2,427) | p > 0.99<br>(V = 2,120,289.5; n = 3,089) |
| One-sided Wilcoxon signed-rank test for AR1 $\tau_K$ > 0 | p = 0.215<br>(V = 1,461,318.5; n = 2,427) | p = 0.712<br>(V = 2,306,715.5; n = 3,089) |

**Predictive information from current meteorological and road-surface conditions and recent antecedent conditions**

To assess incremental predictive information, we compared Model 0, using only a road surface temperature threshold; Model 1, using current road surface temperature and relative humidity; and Model 2, adding recent temperature change and cumulative conditions (Fig. 5).

The area under the precision–recall curve (PR-AUC) increased from 0.015 for Model 0 to 0.042 for Model 1 and 0.085 for Model 2, all above the climatological icing base rate of 0.0088 (Fig. 5b). The Brier skill score, which measures probabilistic predictive performance, also improved from 0.019 for Model 1 to 0.041 for Model 2 (Fig. 5c).

At a common target POD of 0.80, achieved as 0.803 for Model 0, false alarms per station-day decreased from 15.69 for Model 0 to 6.72 for Model 1 and 3.96 for Model 2. The median first-trigger lead times, by contrast, were similar for the three models: 157, 147 and 151 min, respectively (Supplementary Table 7).

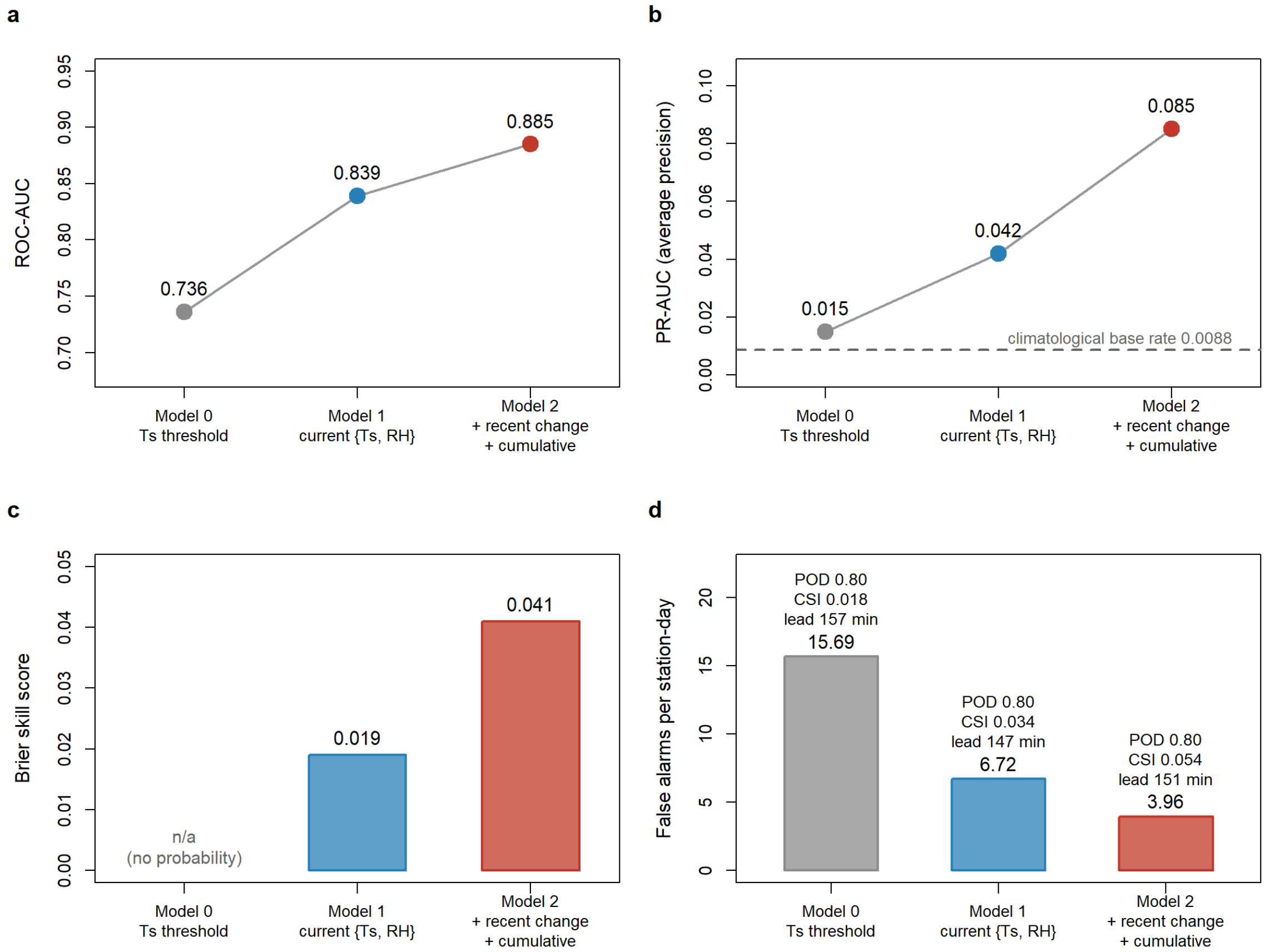

**Fig. 5**

In station-held-out cross-validation, the ROC-AUC of Model 2 decreased from 0.905 at T+1 h to 0.895 at T+2 h, 0.885 at T+3 h, 0.860 at T+6 h and 0.817 at T+12 h. At T+12 h, the 95% confidence interval was 0.805–0.828 (Table 3).

**Table 3**

| **Forecast horizon** | **T+1** | **T+2** | **T+3** | **T+6** | **T+12** |
|---|---|---|---|---|---|
| Station-held-out CV AUC | 0.905 | 0.895 | 0.885 | 0.860 | 0.817 |
| 95% CI | 0.898–0.913 | 0.887–0.902 | 0.877–0.894 | 0.850–0.869 | 0.805–0.828 |

At T+6 h and POD 0.80, median first-trigger lead time was 246 min (about 4 h), with 3.57 false alarms per station-day (Table 4). For the same T+6 h prediction, lowering

POD to 0.70 shortened the median lead time slightly to 229 min and reduced false alarms to 2.14 per station-day. For the T+6 h prediction, a persistence condition requiring threshold exceedance at two consecutive evaluation times (k = 2) gave 1.47 false alarms per station-day at an operating point with POD 0.55. Requiring persistence further reduced false alarms but also lowered icing detection.

**Table 4**

| **Setting** | **POD** | **First-trigger lead (median)** | **False alarms per station-day** |
|---|---|---|---|
| T+6 | 0.80 | **246 min (≈4 h)** | **3.57** |
| T+6 | 0.70 | 229 min | **2.14** |
| T+3 | 0.80 | 151 min | **3.96** |
| T+6, persistence k=2 | **0.55** | 208 min | **1.47** |

We next evaluated whether surrounding meteorological fields provide additional predictive information beyond the local observations and antecedent conditions available at individual RWIS stations.

## Additional predictive information from surrounding meteorological fields

Adding features describing the thermal and moisture state and recent changes in surrounding meteorological fields consistently improved discrimination beyond that obtained from local RWIS observations and recent antecedent conditions at all forecast horizons (Fig. 6a). In station-held-out evaluation, the ROC-AUC of the model with surrounding-field features was 0.030–0.032 higher than that of the model using RWIS information alone across T+1 to T+12 h.

The additional predictive information from surrounding meteorological fields was more evident when false alarms were compared at the same POD (Fig. 6b). With POD fixed at 0.80, false alarms per station-day at T+6 h decreased by 67%, from 3.57 with RWIS information alone to 1.17, and at T+3 h by 62%, from 3.96 to 1.50 (Supplementary Table 7). At non-icing evaluation times with road surface temperature at or below 0 °C, false alarms also decreased by 65–66% at T+6 h and by 60–61% at T+3 h. Median first-trigger lead times shortened from 246 to 203 min at T+6 h and from 151 to 128 min at T+3 h (Supplementary Table 7).

The improvement from surrounding-field features was also consistent across data-splitting schemes in the event–control-time analysis (Fig. 6c). With RWIS information alone, ROC-AUC was 0.948, 0.949 and 0.938 in station-held-out, date-held-out and winter-held-out evaluations, respectively, and adding surrounding-field features raised it to 0.969, 0.971 and 0.961. Because this analysis uses an event–control-time discrimination design that differs from the future icing prediction in Fig. 6a, we evaluated the relative improvement from adding surrounding-field features under each data-splitting scheme rather than the absolute AUC values.

Relative to the model using RWIS information alone, ROC-AUC increased by 0.021, 0.022 and 0.023 in station-held-out, date-held-out and winter-held-out evaluations with same-time fields, by 0.015, 0.017 and 0.017 with fields from 1 h earlier, and by 0.013, 0.014 and 0.011 with fields from 3 h earlier; each increase is the difference between the two AUC values rounded to three decimals.

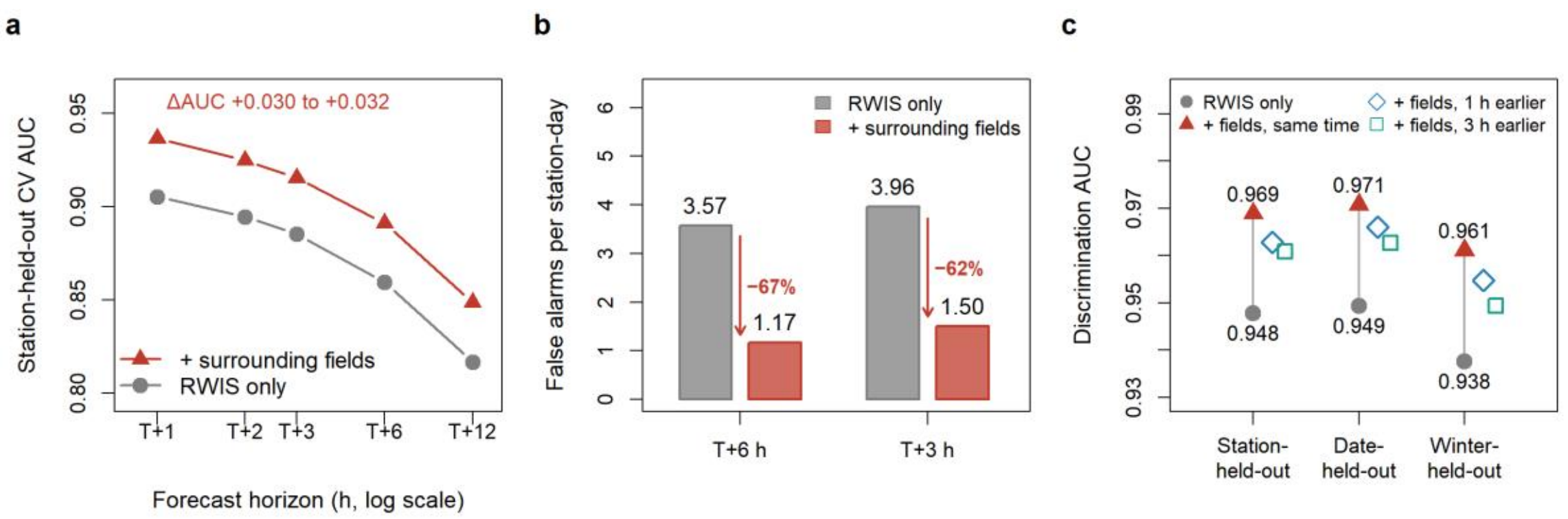

**Fig. 6**

## Discussion

The central result of this study is that CSD-based early warning signals did not consistently appear before icing onset in a large road-weather observation dataset. In 2,427 icing events, the trends of variance and AR1 were not clearly distinguishable from those at randomly selected non-icing control times, and this result held when the analysis window and detrending settings were changed. Moreover, nearly identical results were obtained for 3,091 icing events at 127 independent RWIS stations with no station overlap.

The onset-condition analyses showed weak separation among events with different prior surface states and a largely continuous structure in the observed physical conditions associated with icing onset, rather than a small number of clearly separated pathways.

Cold dose, deposition dose and freeze–thaw accumulation were generally higher before icing than at non-icing control times, indicating an association with antecedent thermal and moisture history rather than road surface temperature at a single time alone. For cold dose, discrimination was highest at short decay time constants of 30 min or less and decreased as longer decay timescales were used, including when cold dose was computed 30–60 min before onset. This empirical timescale should not be interpreted as an intrinsic thermal response time of the road surface; rather, it indicates that recent cooling carried more discriminative information about icing onset than cooling accumulated over longer periods. Besides surface cooling, icing can be influenced by moisture-related conditions, including the presence or supply of liquid water, conditions favourable for water-vapour deposition, antecedent precipitation and snow cover, and freeze–thaw history.

The additional predictive information from surrounding meteorological fields extends this interpretation spatially. The same advantage appeared in event–control evaluations that held out different stations, dates and winters, and it remained, although smaller, in the past-only analysis that excluded same-time objectively analysed fields. These results indicate that information associated with the thermal and moisture state of the surrounding atmosphere contributes to icing prediction beyond the local state history observed at individual RWIS stations.

These results are more consistent with recent meteorological forcing changing the road surface's thermal and moisture state towards icing than with a canonical CSD-type transition involving gradual loss of internal stability near a critical point. Here, meteorologically forced threshold crossing means movement into icing-favourable conditions formed jointly by road surface temperature, moisture presence and potential supply, atmospheric water vapour and their temporal changes, rather than crossing a single fixed threshold.

CSD-based early warning signals reflect changes in recovery dynamics associated with declining internal stability, whereas the predictive information identified here was contained in current meteorological and road-surface conditions, recent antecedent conditions and surrounding meteorological fields.

The stepwise addition of information illustrates this difference. Using current road surface temperature and relative humidity together improved discrimination substantially over a simple road surface temperature threshold, and adding the recent change in road surface temperature and cumulative antecedent conditions, expressed as cold dose, deposition dose and freeze–thaw accumulation, improved it further. Cumulative antecedent conditions alone did not discriminate better than the current state, but they improved predictive performance when added to the current state and recent change. These cumulative indicators therefore functioned not as independent icing thresholds but as complementary predictors that represented recent state history.

The reduction in false alarms from adding surrounding-field features is particularly important in this respect. A low road surface temperature at a station alone was insufficient to distinguish icing from non-icing, but adding the thermal and moisture conditions of the surrounding atmosphere reduced false alarms at the same POD.

These results support combining current meteorological and road-surface conditions, recent changes, cumulative antecedent conditions and surrounding meteorological fields for road-icing prediction, rather than relying on a single road surface temperature threshold or long-term cumulative indicators. Because operational thresholds produced clear trade-offs among POD, lead time and false alarms, alert conditions in operational warning systems need to be set according to operational objectives and acceptable false-alarm levels rather than by maximizing a single performance metric.

Beyond the specific phenomenon of road icing, this study identifies considerations for applying early warning signals of critical transitions to real observations. An observed abrupt change in state alone does not justify assuming that CSD precedes the transition. When state changes are driven more directly by changing external forcing without a progressive loss of stability near a bifurcation, increases in variance or AR1 are not necessarily expected to provide reliable early warning.

Assessing whether a transition can be anticipated therefore requires asking, beyond whether CSD-based early warning signals are present, which processes are involved in the transition and where the predictive information associated with those processes resides.

Operational practice already reflects this distinction. Frost warnings are commonly based on predictions from external conditions such as air temperature and humidity [32], and rainfall-threshold approaches have likewise been used for urban flood warning [33].

Our results do not imply that CSD is generally absent or that CSD theory itself is limited. Rather, they show the importance of distinguishing transitions for which CSD is theoretically expected from cases in which state change is driven primarily by changing external forcing without the progressive loss of stability required for CSD. From this perspective, high-frequency environmental observation networks can provide an empirical basis for testing theoretically proposed early warning concepts in real natural and environmental systems and for evaluating which kinds of advance information are useful depending on the dynamical characteristics of a transition.

Our results should be interpreted in light of several limitations. First, the icing state was defined by the ICE category classified by the RWIS surface-state sensor, which covers various forms of icing on roads and does not refer only to transparent black ice. The high agreement with ice-film thickness (IFT) supports consistency between different sensor-derived variables, but because IFT is also produced by the same RWIS, this is not a validation against fully independent ground truth. Independent validation using cameras or direct observations of road-surface ice is therefore needed.

Second, the variance and AR1 used here are statistical early warning signals common in CSD theory, but RWIS observations do not directly measure the internal stability of the road-surface system or its recovery dynamics after perturbations. Our results therefore show that the CSD-based early warning signals tested did not consistently appear before road icing; they do not rule out every form of nonlinear dynamics or stability change that may exist in the road-surface system. Direct observations of road-surface energy and water budgets, subsurface heat conduction and atmosphere–surface heat exchange would help distinguish meteorological forcing from the road surface’s dynamic response.

Third, the independent network shared no stations with the primary network. It replicated the lack of consistent CSD signals, and the local prediction model retained predictive skill there without retraining. However, it has no on-site meteorological sensors, so air temperature and relative humidity needed for part of the quality control and for the prediction-model inputs were taken from 500 m objectively analysed fields. Future work requires prospective validation using fully independent meteorological observations in a real-time operational setting.

Finally, this study was conducted for expressways and winter climate conditions in South Korea. In regions with different pavement materials, de-icing practices, traffic volumes, radiative environments or precipitation types, the thermal and moisture processes leading to icing and the effective timescales may differ. Validation across climatic and road environments is therefore needed to assess whether the absence of consistent CSD-based early warning signals and the predictive information from recent and surrounding conditions are reproduced elsewhere.

## Conclusions

In 2,427 icing-onset events from a 121-station network, variance and lag-1 autocorrelation did not increase consistently before icing, and this result was replicated at 127 independent stations. Current conditions, recent changes and cumulative antecedent conditions carried predictive information that was retained when the model was applied to the independent network without retraining. Surrounding meteorological fields further improved prediction and reduced false alarms at the same POD.

The lack of consistent CSD-based early warning signals does not imply that icing is unpredictable. Even when no useful early warning signal emerges from recovery dynamics associated with declining internal stability, predictive information may still reside in current conditions, recent changes, antecedent state history and external forcing. CSD should therefore not be assumed from the mere existence of a state transition; instead, the physical processes governing the transition and the sources of predictive information need to be evaluated separately.

## Methods

### RWIS network and analysis period

We used 1-min observations from the fixed road weather information system (RWIS) operated along expressways by the Korea Meteorological Administration (KMA). The analysis covered 121 RWIS stations on 12 expressway routes, spanning 34.8–37.8°N and 126.5–129.3°E and including the major expressway sections of South Korea (Fig. 1a). Station elevations range from 7 to 767 m, with a median of 95 m. The stations comprise 20 core stations equipped with 12 types of meteorological and road-surface sensors and 101 basic stations with 5 types of sensors. The network was expanded in stages during the study period; by installation year, it included 13 stations installed in 2022, 15 in 2023, 58 in 2024 and 35 in 2025.

The analysis period was set to November–March, when road icing mainly occurs, and covered three winters: 2023/24, 2024/25 and 2025/26. The 2022/23 winter was excluded because observations began on 29 December 2022, leaving no data for more than half of the analysis period, and only 2–6 stations were operating, depending on the period. A sensitivity analysis of including the 2022/23 winter is presented in Supplementary Table 5.

To check the spatial correspondence between each RWIS station and its expressway section, we examined the distance between each station and the nearest road link of its recorded expressway route. Three stations located 500 m or more from the nearest road link (9.87 km, 2.88 km and 652 m) were excluded for poor spatial correspondence with their expressway sections. After this data selection and spatial check, the final dataset comprised 121 stations and 48,117,793 observation minutes, with a median of 416,137 minutes per station.

**Observed variables and treatment of missing data**

From the RWIS observations, road surface temperature ($T_s$; raw field TR1), surface state (SFS) and ice-film thickness (IFT) were used as variables describing the thermal and moisture state of the road surface and the presence of icing. The meteorological variables were air temperature ($T_a$; TA), relative humidity (RH; HM), precipitation occurrence (RN05_YN), wind speed (WS), wind direction (WD) and air pressure (PA). SFS is a categorical variable derived from sensor measurements by the manufacturer's classification algorithm, and IFT is a sensor-derived variable representing the thickness of the ice layer on the road surface.

Data availability was very high for the road-surface variables. The mean missing rates across stations were 0.00% for road surface temperature, 0.03% for surface state and 0.10% for ice-film thickness, and the mean missing rate of the meteorological variables (air temperature, relative humidity, precipitation occurrence, wind speed, wind direction and air pressure) was 3.85%. Road surface temperature and surface state, the main variables for analysing icing onset and the preceding temporal changes, were therefore available for most of the analysis period.

Missing values were not interpolated in the analyses of icing events and CSD indicators; gaps within events were filled by linear interpolation only for the supplementary trajectory clustering described below. This is because the CSD test, one of the main analyses, relies on changes in the variance and autocorrelation of pre-icing time series, and interpolating gaps could reduce the variance or artificially alter the temporal autocorrelation structure. Where the 1-min continuity was broken, the data were divided into separate continuous segments for event extraction. In the CSD analysis the missing minutes were left missing within fixed analysis windows, and a window was used only when it met the minimum data availability required. Observation times were defined from the datetime field (YYYYMMDDHHMM) of the raw data.

**Icing-event definition and quality control**

To identify icing events consistently, the RWIS surface state (SFS) was reclassified into four categories. Raw state code 1 was assigned to DRY, codes 2 and 11–14 to WET, code 3 to SNOW, and codes 4 (Ice) and 15 (Snow/Ice) to ICE. Snow/Ice is a mixed state in which snow and ice coexist, but it was included in the ICE category because ice is present on the road surface. This classification separates the surface states before icing onset into DRY, WET and SNOW while combining the states in which ice is actually present into a single ICE category.

The icing-onset time t0 was defined as the first time the surface state changed from DRY, WET or SNOW to ICE. To avoid counting temporary non-icing observations within the same icing process as separate events, consecutive ICE periods separated by non-icing intervals shorter than 60 min were first merged. To avoid misidentifying momentary sensor fluctuations as independent icing events, a merged period was accepted as an icing event only when it contained at least 10 min of ICE observations in total. An icing-onset event in this study is therefore not every time ICE was observed but an event in which the surface changed from a non-icing state to a persistent icing state.

Four quality-control steps were applied to ensure the physical consistency of the data and the reliability of the icing events. The physical-range check was applied in every analysis; the temperature-difference and spike masks were applied when icing events were extracted, and the spike mask also in the independent-network analysis; and the onset-temperature criterion was applied to the event list. First, only data within physical ranges were used: −40 to 40 °C for road surface temperature and air temperature and 0–100% for relative humidity. Second, data with a difference between road surface temperature and air temperature exceeding 30 °C were excluded as likely sensor errors. In the raw data for the analysis period, the maximum difference between the two temperatures was 45.7 °C. In the independent observation network the maximum difference was 24.4 °C, so this criterion excluded no observation there and was instead varied as a sensitivity test. Third, single-time jumps in which road surface temperature changed by 10 °C or more within 1 min and immediately returned were identified as spikes and removed. Finally, icing events with road surface temperature above 5 °C at onset were excluded as having low physical consistency with the meteorological and road-surface conditions. Quality control removed 649 minutes in total (642 minutes violating the temperature-difference criterion and 7 spike minutes), less than 0.002% of the 48,117,793 observation minutes. Separately, 10 candidate icing events were excluded by the onset road surface temperature criterion ($T_s > 5$ °C).

In total, 2,427 icing-onset events were identified at 119 of the 121 stations; no icing onset satisfying the definition was observed at the remaining 2 stations during the analysis period. The surface state immediately before icing was WET for 1,875 events, SNOW for 380 and DRY for 172, so events starting from WET accounted for 77.3% of the total. The number of icing events and their main characteristics by winter are given in Table 1 and Fig. 1b.

Because SFS is a category assigned by the manufacturer's algorithm from sensor signals, its use as the basis for icing events requires confirmation with other observational information. When SFS was compared with IFT at onset, IFT > 0 was observed simultaneously for 94.4% of all events. Because IFT is also derived from sensors installed at the same RWIS, it was not regarded as independent ground truth but was used as auxiliary evidence of cross-sensor consistency for SFS-based icing events.

## Independent observation network

To evaluate whether the results from the primary observation network depended on particular stations or network composition, we used separate RWIS observations that were not used for analysis or model development as an independent observation network. The independent observation network consists of 127 stations installed mainly on road sections where icing occurs frequently, and none of them overlaps spatially with the 121 stations of the primary observation network. The two networks also form separate sets in the KMA station-type codes (Fig. 1a).

Because the independent observation network was built for icing-prone sections, its icing frequency differs from that of the primary observation network. In the 1-min surface-state data for the analysis period, ICE accounted for 1.00% of the time in the independent observation network, about 2.0 times the 0.49% in the primary observation network. The independent observation network was therefore used not simply to add stations but to evaluate the independent replication of the main results and the performance of the road-icing prediction model under a different icing frequency and network composition.

Data from the independent observation network were not used in the analysis of conditions at icing onset, the choice of CSD analysis settings, the determination of timescales for cumulative antecedent conditions or the development of the road-icing prediction model for the primary observation network. After the analysis of the primary observation network was completed, 3,091 icing-onset events were identified independently by applying the same icing-event definition and analysis procedure. With these events, we evaluated the replication of CSD-based early warning signal results and applied the road-icing prediction model trained on the primary observation network to the independent observation network without retraining to evaluate its predictive performance. The independent replication of the CSD results and the performance of the road-icing prediction model were thus evaluated in a different observation network.

**Analysis framework and definition of physical indicators of icing-favourable conditions**

To characterize the state transition of road icing and the predictive information before icing onset, all icing events were aligned to the icing-onset time ($t_0$), and changes in meteorological and road-surface conditions before icing onset were analysed step by step. We first (i) evaluated whether conditions at icing onset form distinguishable pathways and (ii) analysed cumulative antecedent conditions and the decay timescale before icing onset. We then (iii) tested whether critical slowing down appears before icing onset and (iv) evaluated the independent replication of the main results in the independent observation network. Finally, (v) we compared the predictive information about future icing onset provided by current and antecedent meteorological and road-surface conditions and by surrounding meteorological fields.

To represent icing-favourable physical conditions, we defined a liquid-freezing margin ($M_{\mathrm{liq}}$) and a deposition margin ($M_{\mathrm{dep}}$), distinguishing cases with liquid water on the road surface from cases in which atmospheric water vapour can deposit directly onto the surface. For conditions with liquid water, the liquid-freezing margin was defined from the difference between road surface temperature and the freezing point (0 °C) (equation (1)). The liquid-freezing margin is thus a thermal indicator of how far road surface temperature is below freezing, assuming that liquid water is present.

$$M_{\mathrm{liq}} = T_s - 0\ ^{\circ}\mathrm{C} \tag{1}$$

The potential for direct deposition of water vapour was described by the relationship between road surface temperature ($T_s$) and frost-point temperature ($T_f$). The frost-point temperature was calculated from air temperature and relative humidity (equation (2)), and the deposition margin ($M_{\mathrm{dep}}$) was defined (equation (3)).

$$T_f = \frac{b_i l}{a_i - l}, \quad l = \ln\left(\frac{e}{6.112}\right), \quad e = \frac{\mathrm{RH}}{100} \cdot 6.112\ \exp\left(\frac{a_w T_a}{b_w + T_a}\right) \tag{2}$$

where $a_w = 17.62, \quad b_w = 243.12\ ^{\circ}\mathrm{C}, \quad a_i = 22.46, \quad b_i = 272.62\ ^{\circ}\mathrm{C}$ are the Magnus coefficients over water and ice, $e$ is the water-vapour pressure (hPa) and $T_a$ is in °C.

$$M_{\mathrm{dep}} = T_s - T_f \tag{3}$$

The deposition margin ($M_{\mathrm{dep}}$) describes the relationship between road surface temperature ($T_s$) and the frost-point temperature of the surrounding air ($T_f$) and was used as an indicator of conditions favourable for water-vapour deposition on the road surface. The two margins thus separately represent thermal conditions favourable for the freezing of liquid water and conditions favourable for water-vapour deposition, and neither was used as a threshold that directly determines whether icing occurs.

The two margins $M_{\mathrm{liq}}$ and $M_{\mathrm{dep}}$ were used as common physical-condition indicators in the subsequent analyses. The liquid-freezing margin ($M_{\mathrm{liq}}$) was used in the principal component analysis of conditions at icing onset, and the deposition margin ($M_{\mathrm{dep}}$) was used as an input for computing cumulative antecedent moisture conditions and as an alternative state variable in the CSD test. This design separates thermal conditions and moisture-supply conditions relevant to icing for use in the subsequent state-transition analyses, instead of defining icing by a single temperature threshold.

**Structure and continuity of conditions at icing onset**

We evaluated from three complementary perspectives whether the processes leading to icing form distinguishable pathways according to prior surface state or whether events starting from different antecedent conditions show continuous variation.

First, we evaluated how much of the difference in conditions at icing onset is explained by the surface state immediately before icing (DRY, WET or SNOW). For road surface temperature at onset, relative humidity at onset and the 24-h range of road surface temperature before onset, the proportion of variance explained by prior surface state was calculated as the effect size ($\eta^2$) (equation (4)). The liquid-freezing margin equals road surface temperature at onset by equation (1) and was therefore not computed separately; the effect sizes were calculated for the 2,349 events for which physical diagnostics over the 24 h before icing were available.

$$\eta^2 = \frac{SS_{\mathrm{between}}}{SS_{\mathrm{total}}} \tag{4}$$

Here, $SS_{\mathrm{between}}$ is the between-group sum of squares across prior surface states and $SS_{\mathrm{total}}$ is the total sum of squares. $\eta^2$ was used to quantify the proportion of the total variation in conditions at icing onset attributable to prior surface state rather than to assess statistical significance.

Second, we evaluated whether cooling trajectories before icing onset contain information that distinguishes prior surface states. For each icing event, a cooling trajectory was constructed from road surface temperature, deposition margin and relative humidity at 14 times during the 24 h before onset (42 features), and prior surface state (DRY, WET or SNOW) was classified with multiclass LightGBM. The 2,326 events without missing values were randomly split into 70% training data and 30% test data. Classification performance was evaluated by overall accuracy and by balanced accuracy, which accounts for unequal group sizes, and was compared with a simple baseline that always selects the most frequent class. This analysis evaluated how much information distinguishing prior surface states is contained in the temporal cooling process before icing onset.

Third, we evaluated whether the physical characteristics of icing events form a discrete structure with separate groups. Seven physical variables describing icing onset and the preceding processes, namely the preceding 24-h range of road surface temperature, the time of day of onset, the fraction of minutes with observed precipitation in the preceding 24 h, the mean night-time (22–07 h) difference between road surface temperature and air temperature, the logarithm of cold dose, relative humidity at onset and the liquid-freezing margin, were standardized for the 2,342 events without missing values. Principal component analysis (PCA) was then performed, and Hartigan's dip test was applied to the distribution of the first principal component (PC1) [34]. The null hypothesis of this test is that PC1 follows a unimodal distribution. We evaluated whether the observations departed from unimodality and showed significant multimodality without specifying the number of clusters in advance. Correlations between PC1 and individual physical variables were also analysed to interpret the physical characteristics of the variation along PC1.

To complement the unimodality test, a k-means cluster analysis was also performed on the 24-h trajectories of road surface temperature, deposition margin and relative humidity before icing (42 features). The 2,349 events with at least 70% observation availability were used, and gaps within events were filled by linear interpolation. For each number of clusters ($k$ = 2–6), the mean silhouette coefficient was calculated with 25 initializations to select the optimal number of clusters, and the final clusters were formed with 50 initializations at the selected number. The separation of the selected cluster structure was evaluated with the silhouette coefficient, and the correspondence between the derived clusters and the surface state immediately before icing (DRY, WET or SNOW) was quantified with the adjusted Rand index (ARI).

These analyses did not presuppose any particular pathway or cluster structure; the association between prior surface state and conditions at icing onset, the information on prior surface state contained in cooling trajectories, and the distribution and cluster separation of multivariate physical variables were evaluated as complementary lines of evidence. Together, they tested whether observed icing events follow pathways clearly distinguished by prior surface state or are more consistent with a continuous structure encompassing diverse antecedent conditions.

**Accumulation of antecedent conditions and their decay timescale**

To evaluate the relationship between icing onset and the temporal accumulation of antecedent conditions, antecedent cooling, conditions allowing moisture supply and freeze–thaw history were represented as cumulative antecedent conditions. Each antecedent condition was accumulated with a first-order leaky integrator in which the influence of past inputs decays exponentially with time. The continuous-time form of the cumulative state S and its discrete form for 1-min observations are defined as follows (equation (5)).

$$\frac{dS}{dt} = -\frac{S}{\tau_d} + x(t) \quad \Rightarrow \quad S_t \approx S_{t-1}\, e^{-\Delta t/\tau_d} + x_t\, \Delta t\, 1[\Delta t \leq 15] \tag{5}$$

Here, $S_t$ is the cumulative value at time $t$, $x_t$ is the input at that time, $\Delta t$ is the interval between consecutive observations (min) and $\tau_d$ is the decay time constant, which sets how quickly the influence of past inputs declines. A smaller $\tau_d$ gives more weight to recent conditions, and a larger value retains the influence of earlier conditions in the cumulative value for longer. The indicator function $1[\Delta t \leq 15]$ adds new input only when the observation interval is 15 min or less. For data gaps longer than 15 min, no new input was added and only temporal decay was applied to the existing cumulative value. To distinguish it from Kendall's $\tau$ ($\tau_K$) used in the CSD trend tests, the decay time constant is denoted $\tau_d$.

The decay term of the discrete recursion matches the continuous-time solution exactly, and the input term was computed with a first-order approximation assuming that $x$ is constant within each observation interval. The resulting approximation error is proportional to $\Delta t/\tau_d$: for $\Delta t = 1\ \mathrm{min}$, it was 0.8% at $\tau_d = 1\ \mathrm{h}$ and 0.03% at $\tau_d = 1{,}440\ \mathrm{min}$. Because 99.995% of consecutive observation intervals in our data were 1 min, the proportionality factor of this approximation was effectively the same for events and control times. When cold dose and deposition dose were recomputed with the exact solution assuming constant *x* within each observation interval, the icing–non-icing discrimination ROC-AUC was unchanged at both $\tau_\mathrm{d}$ = 1 h and 1,440 min.

The cumulative antecedent-condition indicators consist of cold dose, deposition dose and freeze–thaw accumulation (FT) (equation (6)).

$$\text{cold dose: } x_{\mathrm{cold}} = \max(0, -T_s)$$

$$\text{deposition dose: } x_{\mathrm{dep}} = \max\left(0, -M_{\mathrm{dep}}\right)$$

$$\text{freeze–thaw accumulation: } \mathrm{FT}_t = \mathrm{FT}_{t-1}\, e^{-\Delta t/\tau_d} + 1\left[\operatorname{sgn} T_{s,t} \neq \operatorname{sgn} T_{s,t-1}\right] 1[\Delta t \leq 15] \qquad (6)$$

Cold dose represents the intensity and persistence of antecedent cooling by accumulating, with decay, how far and for how long road surface temperature fell to or below 0 °C. Deposition dose represents the intensity and persistence of antecedent conditions favourable for water-vapour deposition by accumulating, with decay, the magnitude and duration of negative values of the deposition margin defined above ($M_{\mathrm{dep}}$). Freeze–thaw phenomena have been treated as recurrent thermal processes associated with temperature changes around the freezing point [35]. To quantify recent freeze–thaw history, a change in the sign of road surface temperature between two consecutive observations was counted as one 0 °C crossing (observations with road surface temperature of exactly 0 °C were excluded from the sign determination), and the same exponential decay was applied to form the freeze–thaw accumulation. In equation (6), the sign of the preceding observation is that of the most recent observation with road surface temperature other than 0 °C, so that a sign change separated by observations at exactly 0 °C is counted once and a return to the same sign is not counted. The sign is cleared whenever the interval since the previous observation exceeds 15 min, and it is re-established at the next observation that has an interval of 15 min or less and an observed road surface temperature other than 0 °C; no crossing is counted until the sign has been re-established. The cumulative indicators were computed with a default $\tau_d$ = 1,440 min, which was not assumed to be an optimal or intrinsic physical timescale. To evaluate the effective timescale of antecedent conditions empirically, a sensitivity analysis was performed in which $\tau_d$ was varied from 1 min to 72 h. The cumulative indicators were computed at the icing-onset time and at 30 and 60 min before onset, to check whether the same pattern appeared in antecedent cumulative values that exclude road surface temperature at onset. Differences in AUC between $\tau_{\mathrm{d}}$ values were evaluated with a paired station-cluster bootstrap (B = 500) that resampled the same events and control times by station, and AUCs stratified by 3-h time-of-day bins were also compared to account for the concentration of icing onset at particular times of day. The icing–non-icing discrimination of the cumulative indicators computed for each $\tau_d$ was compared using the area under the receiver operating characteristic curve (AUC), and the time constant with the highest discrimination, $\tau_d^*$, was identified. $\tau_d^*$ was not interpreted as an intrinsic physical time constant of the road surface but was defined as the empirical timescale at which the accumulation of antecedent conditions provides the most predictive information about icing in these observations.

The discrimination of the cumulative indicators was evaluated by comparing icing events with non-icing control times. For each icing event, five control times were drawn at random from non-icing times at the same station, and only times at least 12 h away from every icing-onset time were used as control-time candidates to minimize the influence of icing events. The same accumulation procedure was applied to icing events and control times. This design evaluated the discriminative information of cumulative antecedent conditions related to icing onset, and their effective timescale, separately from instantaneous icing-favourable physical conditions.

### Testing critical slowing down-based early warning signals

We tested whether critical slowing down, characterized by a decline in the internal stability of the road-surface system and the resulting slowing of recovery, appears before icing onset. CSD is the progressive slowing of recovery to equilibrium after small perturbations as a system approaches a critical point at which it loses internal stability, and in observed time series it can typically appear as increases in variance and lag-1 autocorrelation (AR1). We aligned the pre-onset time series of each icing event to the icing-onset time $t_0$ and, with road surface temperature as the primary state variable, evaluated whether such CSD-based early warning signals appear before icing onset.

***Primary CSD indicators and trend tests***

To reduce the influence of low-frequency variability and long-term cooling trends in road surface temperature on the CSD indicators, a low-frequency trend estimated by smoothing was removed from the observed road surface temperature. Rolling windows were applied to the detrended residual series r over the pre-onset period of each icing event to compute the primary CSD indicators, variance and AR1, continuously. The residual series and the temporal trend of each CSD indicator are defined as follows (equation (7)).

$$r_t = T_{s,t} - \hat{T}_{s,t}$$
$$\mathrm{trend} = \tau_{\mathrm{K}}(I, t), \quad I \in \{\mathrm{Var}(r),\ \mathrm{AR1}(r)\} \tag{7}$$

Here, $T_{s,t}$ is road surface temperature at time t, $\hat{T}_{s,t}$ is the low-frequency trend estimated by smoothing and $r_t$ is the detrended residual. Whether each CSD indicator increased or decreased towards the icing-onset time was evaluated with Kendall's rank correlation coefficient with time, $\tau_{\mathrm{K}}$. $\tau_{\mathrm{K}} > 0$ indicates that the CSD indicator tends to increase on approach to icing onset, and $\tau_{\mathrm{K}} < 0$ indicates a decreasing tendency. Detrending, rolling-window computation of variance and AR1, and evaluation of temporal trends with Kendall's rank correlation followed procedures widely used in CSD-based early warning signal analysis [36]. To distinguish it from the decay time constant $\tau_{\mathrm{d}}$ of the cumulative antecedent conditions, Kendall's rank correlation coefficient is denoted $\tau_{\mathrm{K}}$. In the primary analysis, the 720 min immediately before icing onset was used as the analysis window. The trend was estimated with a Gaussian smoother with a bandwidth of 60 min, and variance and AR1 were computed in 360-min rolling windows moved in 30-min steps. Analysis windows with less than 80% non-missing data were not used.

The theoretical basis for using variance and AR1 as CSD indicators follows from a local first-order stochastic differential equation approximation of small fluctuations around a stable state. Approximating small fluctuations of road surface temperature as a stochastic process around a time-varying local equilibrium $\mu(t)$ gives the following expression (equation (8)).

$$dT_s = -\lambda\big(T_s - \mu(t)\big)dt + \sigma dW_t \tag{8}$$

Here, $\lambda$ is the recovery rate towards equilibrium, $\mu(t)$ is the time-varying local equilibrium road surface temperature, $\sigma$ is the magnitude of stochastic fluctuations and $dW_t$ is the increment of a Wiener process. This expression was not intended to assume that the full dynamics of road surface temperature follow a stochastic differential equation; it was used as a local approximation to express the theoretical relationship between changes in the recovery rate and in variance and AR1 under CSD.

As internal stability declines on approach to a critical point, the recovery rate $\lambda$ decreases and the relaxation time $\lambda^{-1}$ increases. In theory, this slowing of recovery can appear as an increase in variance and an approach of AR1 towards 1. A consistent increase in variance and AR1 before icing onset was therefore used as the main criterion for the presence of statistical early warning signals consistent with CSD. The variance in equation (8) is the variance of deviations from the local equilibrium, and the detrended residual $r_t$ in equation (7) was used as the empirical counterpart of these deviations. When $\mu$ in equation (8) is constant, the variance of deviations from the local equilibrium is $\sigma^2/(2\lambda)$, and the lag-1 autocorrelation at time step $\Delta t$ is $\exp(-\lambda\Delta t)$.

***Robustness tests across analysis settings and state variables***

To evaluate whether the CSD test results depended on particular settings of the rolling window or detrending, the same test was repeated for nine analysis settings combining three analysis-window lengths (6, 12 and 24 h; rolling window half the analysis window) and three detrending bandwidths (30, 60 and 120 min). For each setting, the temporal changes in variance and $\mathrm{AR1}$ and Kendall's $\tau_{\mathrm{K}}$ were recomputed to evaluate whether the direction and magnitude of the signal were consistent across analysis settings. Control times were redrawn for these robustness tests, so the values for the primary setting differ slightly from those in Table 2 (0.501 versus 0.488 for the variance trend).

To prevent the CSD assessment from depending on the choice of variance and $\mathrm{AR1}$ as indicators, the same trend test was performed for six CSD indicators: variance, AR1, skewness, kurtosis, spectral reddening and autocorrelation decay time. For each indicator, the proportion of events with an increase towards icing onset and the distribution of $\tau_K$ were compared to determine whether any change was specific to a particular indicator or common to multiple CSD indicators.

To evaluate whether temporal changes observed before icing events could arise from general variability in road surface temperature, such as seasonal or diurnal variation, two types of non-icing controls were used. The first consisted of random control times drawn from times without icing, and the second of control times matched to the icing events for time of day within ±1 h (time-matched controls). At most two control times were drawn per icing event at the same station, and every control time was at least 24 h away from any icing onset; the controls were not matched by month or winter. The same preprocessing, rolling-window and trend-test procedures were applied to icing events and to each control set to evaluate whether changes in CSD indicators were specific to icing onset.

Road surface temperature was the state variable in the primary analysis, but the same CSD analysis was applied to alternative state variables to check whether the results depended on this choice. These included the deposition margin defined above, air temperature, and sensor-derived variables related to road-surface friction and water film at stations where these data were available. This analysis evaluated whether similar CSD signals appeared before icing in variables related to moisture and road-surface state as well as to thermal state.

### ***Positive control using synthetic data***

To distinguish whether weak or absent CSD signals in the observations reflected a genuine absence of precursors or an insufficient detection capability of the analysis procedure, a positive control with synthetic time series was performed. The synthetic data were generated from a linear stochastic process whose recovery rate declines towards the end of the window following fold scaling, to generate the increases in variance and AR1 expected under a declining recovery rate. With these series, the same detrending, rolling-window, variance and AR1 computation and Kendall trend-test procedures as for the RWIS data were applied. Control series with a fixed recovery rate ($\lambda$ = 0.05 per minute, AR1 = 0.95), two per transition series, were generated in the same way.

This analysis was not intended to map the synthetic results directly onto real road-icing dynamics; it served as a procedural check of whether the analysis procedure could detect the early warning signals expected under synthetic conditions constructed to contain CSD.

### *Assessment of recovery rate using relaxation time*

To complement the trend analysis of variance and AR1, the recovery speed before icing onset was assessed indirectly through the recovery rate $\lambda$ in equation (8). Because the time required for the system to return to equilibrium lengthens as $\lambda$ decreases in equation (8), the relaxation time, the characteristic time of recovery, was defined as $\lambda^{-1}$. A larger $\lambda^{-1}$ indicates slower recovery to the original state after a perturbation. If CSD preceded icing, $\lambda^{-1}$ would therefore be expected to increase on approach to icing onset. However, interpreting $\lambda^{-1}$ as a relaxation time requires the autocorrelation of the residuals to decay exponentially with lag, so the shape of the autocorrelation decay and the sensitivity of the $\lambda^{-1}$ estimates to the detrending bandwidth (30, 60, 120 and 240 min) were also evaluated. $\lambda^{-1}$ was estimated from AR1 computed in the last rolling window of the analysis window as $\lambda^{-1} = -\Delta t/\ln(\mathrm{AR1})$ ($\Delta t$ = 1 min, $0 < \mathrm{AR1} < 1$).

The $\lambda^{-1}$ estimated for icing events was compared with values at non-icing control times to evaluate whether the relaxation time lengthened before icing onset. This analysis complemented the assessment of CSD from changes in variance and AR1 alone and evaluated whether the decline in recovery rate, the core dynamical property of CSD in equation (8), appeared consistently in the observations.

In these CSD analyses, the presence of CSD was not judged from changes in a single indicator or under a particular analysis setting alone. Trend tests of the primary CSD indicators (variance and AR1), robustness tests across analysis settings, comparisons with non-icing control times, analyses of alternative state variables, a positive control with synthetic data and relaxation-time analysis were applied as complementary tests of whether CSD-based early warning signals consistent with slowing recovery appeared repeatedly before icing onset.

### *Independent replication of CSD results in a separate observation network*

To evaluate whether the CSD test results from the primary observation network depended on particular stations or network composition, independent replication was evaluated in a separate RWIS network that was not used for analysis or for choosing method settings. The independent observation network consists of the 127 stations described above, none of which overlaps spatially with the 121 stations of the primary observation network. The two networks are also separated in the station-type codes, so the independent observation network was treated as a set of stations distinct from the primary observation network.

The same icing-event definition and CSD-based early warning signal analysis procedure as for the primary observation network were applied to the independent observation network. The same code and analysis functions used for the primary observation network were applied unchanged to the independent observation network. The main CSD analysis procedures, including the duration and merging criteria for icing events, detrending, rolling windows, computation of variance and AR1, and Kendall's τK trend test, were therefore identical in the two networks.

The two networks differ, however, in the available meteorological observations. Because the independent observation network has no on-site air temperature or relative humidity sensors, air temperature and relative humidity from the 500 m objectively analysed fields were used for these variables.

However, the primary CSD analysis relied on road surface temperature and surface state, so meteorological data from the objectively analysed fields were not used directly to compute CSD indicators. In the CSD analysis, air temperature was used only in the quality-control step that excludes data whose difference between road surface temperature and air temperature exceeds the tolerance. To check whether the results for the independent observation network were sensitive to the source of air temperature data or to the quality-control criterion, the same analysis was repeated with the tolerance for the difference between road surface temperature and air temperature tightened from the default 30 °C to 15 °C and 10 °C. This test evaluated the effect of using air temperature from the objectively analysed fields, and of the choice of quality-control criterion, on the CSD test results.

Because the number of stations in both networks increased in stages during the study period, we also examined whether results pooled over the whole period could be dominated by a particular winter or by stations added at a particular time. For this purpose, the same CSD analysis was repeated with both networks stratified by winter, and the trend characteristics of variance and AR1 and the discrimination between icing and non-icing were compared across winters. This analysis was performed as a sensitivity test of how changes in sample composition caused by the yearly expansion of the networks affected the overall results.

Independent replication was evaluated by comparing, on the same basis, the distributions of Kendall's $\tau_K$ for rolling variance and AR1 and the discrimination between icing and non-icing in the primary and independent observation networks. Confidence intervals were calculated with a station-cluster bootstrap (B = 5,000) to account for correlation among icing events at the same station. Going beyond validation by data splitting within one network, this comparison tested whether the CSD test results from the primary observation network were reproduced in a spatially separate network with a different icing frequency and data composition.

**Road-icing prediction and evaluation of generalization performance**

To evaluate the predictive information that current and antecedent meteorological and road-surface information provides about future icing onset, an observation-based probabilistic prediction model was built. Evaluation times t were set at 30-min intervals during non-icing states, and the probability of a new icing onset within H=1, 2, 3, 6, 12 h after each evaluation time was estimated. The evaluation sample comprised 1,478,456 evaluation times at the 119 of the 121 stations that had eligible icing events and complete inputs. The inputs were current road surface temperature and relative humidity, the change in road surface temperature over the last hour, and the cold dose, deposition dose and freeze–thaw accumulation defined above. All inputs were computed only from information available up to the evaluation time t.

The probability of future icing onset was estimated with LightGBM [37]. This analysis focused not on optimizing a particular prediction model but on evaluating the additional information that different inputs provide for predicting future icing onset. For this purpose, current road surface temperature and relative humidity, the recent change in road surface temperature and cumulative antecedent conditions were added step by step to a simple temperature threshold, and the change in discrimination of icing onset at each step was compared. The direction and relative magnitude of the contribution of individual inputs to model predictions were evaluated with SHAP (SHapley Additive exPlanations) [38]. Detailed hyperparameters and training conditions are given in Supplementary Methods 1.

To reduce overestimation of predictive performance caused by the strong spatial and temporal dependence of RWIS data, generalization performance was evaluated with block cross-validation [39]. Spatial generalization was evaluated with station-held-out five-fold cross-validation, and temporal generalization with winter-held-out cross-validation in which each winter was held out in turn. Performance in the independent observation network was evaluated by applying the model trained on the primary observation network to the independent observation network without retraining. For the independent observation network, air temperature and relative humidity, which are not observed on site, were taken from the 500 m objectively analysed fields.

Discrimination of icing onset was evaluated with ROC-AUC and with PR-AUC, which accounts for the low frequency of icing events [40,41]. Probabilistic predictive performance was evaluated with the Brier score [42], and probability calibration was checked with reliability analysis [43,44]. Operational alert performance was compared using false alarms per station-day at a target POD and the lead time from the first valid alert to icing onset. Detailed evaluation procedures, including the probability calibration method, the treatment of consecutive alerts, the counting of false alarms and cost–loss analysis [45,46], are given in Supplementary Methods 1.

**Additional predictive information from surrounding meteorological fields and assessment of same-time information leakage**

To evaluate whether surrounding meteorological fields provide additional predictive information about future icing onset beyond RWIS station observations, we used the KMA objectively analysed fields with a spatial resolution of 500 m and a temporal resolution of 5 min. Air temperature and dew-point temperature from these fields were matched in space and time to each RWIS station. Features derived from surrounding meteorological fields were designed to represent temporal and spatial changes in cooling and moisture conditions around each RWIS station and consisted of recent temporal changes, 10-km-scale surrounding fields and upwind conditions. Specifically, at the grid cell corresponding to each RWIS station, we used the 30-min changes in air temperature, dew-point temperature and dew-point depression; the deviations of air temperature and dew-point temperature from their means within a 10 km radius; the dew-point depression at grid cells 5, 10 and 20 km upwind along the observed wind direction; and the air temperature difference between the grid cell 10 km upwind and the station grid cell. Details of the spatial feature computation and selection are given in Supplementary Methods 2.

The information provided by features derived from surrounding meteorological fields was first evaluated as discriminative information in an event–control-time analysis using icing-onset events and non-icing control times at the same stations; the same features were then added to the road-icing prediction model described above to evaluate additional predictive information about future icing onset. To reduce overestimation of predictive performance caused by spatial and temporal dependence, station-held-out, date-held-out and winter-held-out block cross-validation were applied. Predictive performance was evaluated by comparing the ROC-AUC of the model using RWIS information alone with that of the model with features derived from surrounding meteorological fields added, and operationally by the false-alarm frequency and lead time at the same target POD.

To evaluate whether predictive performance improved because same-time objectively analysed fields shared information with the current RWIS state, a past-only analysis was performed. With the objectively analysed field at the evaluation time excluded, features derived from surrounding meteorological fields were recomputed with reference to a time 1 h or 3 h before the evaluation time, using only objectively analysed fields up to that time and the observed wind direction at that time. The same evaluation was then repeated. This analysis evaluated how much the additional predictive information from surrounding meteorological fields depended on same-time data and whether it was retained with only the surrounding meteorological fields available before the evaluation time.

**Statistical analyses**

All analyses were performed in R 4.5.1. Discrimination between icing events and non-icing control times, and between evaluation times followed or not followed by icing onset, was quantified with the area under the receiver operating characteristic curve (ROC-AUC) computed from the Mann–Whitney rank statistic and, for the rare-event prediction task, with the area under the precision–recall curve (PR-AUC). Probabilistic predictions were evaluated with the Brier score and the Brier skill score relative to the climatological base rate. Generalization was estimated with station-held-out five-fold, winter-held-out and, for the event–control-time analysis, date-held-out block cross-validation. Confidence intervals (95%) were obtained as the 2.5th and 97.5th percentiles of station-cluster bootstraps that resample stations with replacement (B = 5,000 for the CSD discrimination AUCs in Table 2 and B = 1,000 for the prediction AUCs in Table 3). Differences in AUC between decay time constants were evaluated with a paired station-cluster bootstrap (B = 500) and regarded as indistinguishable when the 95% interval of the difference included zero. Temporal trends in CSD indicators were measured with Kendall's rank correlation coefficient with time, and whether event-level trends were positive was tested with one-sided Wilcoxon signed-rank tests, which treat events as independent; dependence among events at the same station is accounted for in the station-cluster bootstrap intervals of the AUCs. Unimodality of the first principal component was tested with Hartigan's dip test; cluster separation was measured with the mean silhouette coefficient and agreement between clusters and prior surface state with the adjusted Rand index. The share of variance explained by prior surface state was measured with $\eta^2$, and associations with the first principal component with Pearson's correlation coefficient.

## Statements and Declarations

### Acknowledgements

The authors thank the National Institute of Meteorological Sciences of the Korea Meteorological Administration for providing the 1-min road weather information system observations used in this study.

### Funding

This research was supported by Global - Learning & Academic research institution for Master's·PhD students, and Postdocs(G-LAMP) Program of the National Research Foundation of Korea(NRF) grant funded by the Ministry of Education(No. RS-2026-25610465).

### Competing Interests

The authors declare no competing interests.

### Author Contributions

Both authors contributed to the study conception and design. Data curation, implementation of the analysis pipeline, and the statistical and machine-learning experiments were performed by Yun Am Seo. Meteorological interpretation and the objectively-analysed-field spatial pattern analysis were contributed by Chaeyeon Yi. The first draft of the manuscript was written by Yun Am Seo and both authors commented on previous versions. Both authors read and approved the final manuscript.

### Data Availability

The 1-min road weather information system (RWIS) observations are produced by the Korea Meteorological Administration (KMA). Records from 15 December 2025 onwards are publicly available through the KMA API Hub (https://apihub.kma.go.kr);

earlier records, which were provided for this study by the National Institute of Meteorological Sciences of KMA, are available from KMA on request and cannot be redistributed by the authors. The 500 m objectively analysed fields are also produced by KMA and are available from KMA. The aggregated data underlying every figure and table and the CSD indicator values of all icing-onset events and control times in both observation networks are available in the same repository (https://github.com/yunam-seo/road-icing-forced-threshold). Event-level data derived from the RWIS observations are available from the corresponding author on reasonable request, subject to the KMA conditions of use.

## Code Availability

The R code implementing the event definition, cumulative antecedent-condition indicators, continuity of conditions at icing onset, critical-slowing-down analysis, and road-icing prediction procedures, together with the scripts that generate every figure and table, is openly available under the MIT licence at https://github.com/yunam-seo/road-icing-forced-threshold. Analyses were run in R 4.5.1 with the data.table (1.18.4), lightgbm (4.6.0), terra (1.8.54), diptest (0.77.1), cluster (2.1.8.1) and maps (3.4.3) packages.

## Figure legends

**Fig. 1.** Observation networks, icing-onset events and analysis design. (a) The primary observation network (121 RWIS stations) and the independent observation network, with no overlapping stations. (b) Icing-onset events per winter in the two networks. (c) Analysis flow: conditions at icing onset, cumulative antecedent conditions and CSD indicators are examined in 1-min observations, followed by analyses of predictive information from current conditions, antecedent conditions and surrounding meteorological fields. RWIS, road weather information system.

**Fig. 2.** Conditions at icing onset show a largely continuous structure with weak pathway separation. (a) Distributions of road surface temperature at icing onset overlap strongly among prior surface states. (b) Prior surface state accounts for at most 6.4% of the variation in conditions at icing onset ($\eta^2 = 0.064$). (c) Median cooling trajectories by prior surface state (shading, IQR); differences among prior states are largest 6–8 h before onset and narrow towards onset, with road surface temperature remaining lowest throughout for events starting from SNOW. (d) The first principal component of physical features shows no clear multimodality (Hartigan's dip test, p = 0.99). (e) Physical characteristics vary gradually along PC1; events are grouped into PC1 deciles, and points show the within-decile mean of each feature against the median PC1, with each feature rescaled to 0–1 across the decile means. IQR, interquartile range; PC1, first principal component of the physical features.

**Fig. 3.** Cumulative antecedent conditions and sensitivity to decay timescale before road-icing onset. (a–c) Cold dose, deposition dose and freeze–thaw accumulation evaluated at icing onset and at non-icing control times; centre lines, medians; boxes, lower to upper hinges; whiskers, most extreme values within 1.5 times the box height; outliers not shown. (d) Discrimination AUC across decay time constant $\tau_{\mathrm{d}}$ for cold dose computed at onset and 30 and 60 min before onset. (e) Incremental predictive information from current conditions, recent change in road surface temperature, and cumulative antecedent conditions for T+3 h road-icing prediction. In (d), yellow circles mark the highest AUC among the decay timescales for each curve. In (e), the two labelled increments are relative to current conditions and to current conditions with recent change, respectively.

**Fig. 4. Lack of consistent CSD-based early warning signals before road-icing onset.** (a, b) Distributions of Kendall's $\tau_K$ for rolling variance and lag-1 autocorrelation before icing onset and at non-icing control times in the primary observation network. (c) Discrimination AUC across nine window and detrending-bandwidth combinations. (d) Discrimination AUC with 95% station-cluster bootstrap confidence intervals for the primary and independent networks. AUC, area under the ROC curve; CSD, critical slowing down; ROC, receiver operating characteristic.

**Fig. 5.** Predictive information added step by step in the road-icing prediction model (station-held-out cross-validation, T+3 h). Model 0 uses a road surface temperature threshold; Model 1 uses current road surface temperature and relative humidity; and Model 2 additionally includes the recent change in road surface temperature and cumulative antecedent conditions. (a) Discrimination (ROC-AUC). (b) Precision–recall performance under extreme class imbalance; dashed line, climatological base rate of 0.0088. (c) Probabilistic predictive performance (Brier skill score); Model 0 yields no probability. (d) Operational trade-off at the same target POD. ROC-AUC, area under the receiver-operating-characteristic curve; PR-AUC, area under the precision–recall curve (average precision); POD, probability of detection; CSI, critical success index.

**Fig. 6.** Surrounding meteorological fields reduce false alarms at the same probability of detection. (a) Station-held-out cross-validation AUC by forecast horizon with and without surrounding-field features (recent change, 10-km anomaly and upwind conditions derived from 500 m objectively analysed fields). (b) At equal detection (POD 0.80), false alarms per station-day decrease by 67% (T+6) and 62% (T+3); median first-trigger lead times shorten from 246 to 203 min (T+6) and from 151 to 128 min (T+3) (Supplementary Table 7). (c) Event–control discrimination under station-, date- and winter-held-out cross-validation; the AUC improvement decreases but is retained when the surrounding-field features are computed only from fields 1 h or 3 h before the evaluation time. AUC, area under the ROC curve; POD, probability of detection; RWIS, road weather information system.

**Table legends**

**Table 1.** Summary of icing events: number of events, stations with events, prior surface state at onset, cross-sensor consistency based on ice-film thickness (IFT > 0), and distribution by winter.

**Table 2.** Comparison of CSD indicators between the primary observation network (primary set, 121 stations) and the independent observation network (independent set, 127 stations), with no overlapping stations. $\tau_K$, Kendall's rank correlation coefficient of the indicator trend; AR, lag-1 autocorrelation (AR1); V, Wilcoxon signed-rank statistic; AUC, area under the ROC curve; CI, station-cluster bootstrap confidence interval.

**Table 3.** Discrimination of the road-icing prediction model across forecast horizons in station-held-out cross-validation, with station-cluster bootstrap confidence intervals (B = 1,000 station resamples). T+N, forecast horizon of N hours; CV, cross-validation; AUC, area under the ROC curve.

**Table 4. Operational alert performance: median first-trigger lead time and false alarms per station-day (alert rate at non-event evaluation times, normalized to a 24-h day).** The probability threshold is set to achieve the stated POD; the alert condition is met when the predicted probability is at or above this threshold at k consecutive 30-min evaluation times (k = 1 unless stated), and the first valid alert is

the first time this condition is met for an icing event. T+N, forecast horizon of N hours; POD, probability of detection.

**Supplementary Information**

**Early warning of road icing from antecedent meteorological conditions without consistent critical slowing down**

## Supplementary Methods 1. Training, probability calibration and operational alert evaluation of the road-icing prediction model

### 1.1 LightGBM model configuration

The road-icing prediction model was built with LightGBM [1]. The aim of this study is not to achieve maximum predictive performance through hyperparameter optimization but to compare, under consistent conditions, the predictive information that current meteorological and road-surface conditions, recent changes and cumulative antecedent conditions provide about future icing onset. The same model configuration was therefore used for all forecast horizons and validation schemes.

The LightGBM objective was binary, with a learning rate of 0.05, num_leaves of 31, min_data_in_leaf of 100, feature_fraction and bagging_fraction of 0.8 each, and a bagging frequency of 1. The number of boosting iterations was fixed at 300, and verbosity was set to −1. All other parameters were left at the LightGBM defaults. Class weighting for positive events was not used in the primary analysis; weighted training with scale_pos_weight was evaluated only in a separate comparison. For reproducibility, the R random seed was fixed with set.seed(2026), and this seed was used to assign stations to cross-validation folds. No separate LightGBM seed was specified, and the library default was used.

Evaluation times were set at 30-min intervals during non-icing states. This interval reduces excessive overlap among evaluation samples caused by the high temporal autocorrelation of adjacent times in 1-min RWIS data while retaining a sufficient number of evaluation samples. At each evaluation time t, the features were computed only from information available up to that time. The basic features were current road surface temperature and relative humidity, the change in road surface temperature over the last hour, and three indicators of cumulative antecedent conditions: cold dose, deposition dose and freeze–thaw accumulation.

The prediction target was whether a new icing onset occurred within a forecast horizon H of 1, 2, 3, 6 or 12 h after each evaluation time t. The model therefore does not predict the road-surface state at a specific future time; it is a probabilistic prediction model that estimates the probability of a new icing onset within each forecast horizon. H = 3 h was used as the reference forecast horizon for the detailed comparison of operational alert performance.

### 1.2 Rare-event handling and probability calibration

Because icing onset is a rare event, with a positive rate of about 0.9% in the evaluation data, model discrimination was evaluated with ROC-AUC and with the precision–recall AUC (PR-AUC), which assesses the discrimination of rare positive events more directly [2, 3]. To evaluate the relationship between icing detection and false alarms as the probability threshold varies, a performance diagram combining the probability of detection (POD) and false-alarm characteristics was also used [4].

To examine how positive-class weighting affects model discrimination and predicted probabilities in rare-event learning, weighted training with the LightGBM scale_pos_weight parameter was compared with unweighted training. To evaluate the effect of post hoc probability calibration, an unweighted model calibrated with isotonic regression was also compared [5, 6]. Three configurations were therefore compared for probabilistic prediction: a positive-class-weighted model, an unweighted model and an unweighted model calibrated with isotonic regression.

The data for isotonic regression were kept independent of the test data of each cross-validation fold. The training data of each fold were split again 80:20; a LightGBM model was fitted to the 80% subset, and an isotonic calibration function was fitted to the predicted probabilities and observations of the remaining 20%. This inner split was random within the training fold and was not blocked by station or date. The calibration function fitted on the inner 20% was then applied to the test-fold predictions of the model trained on the full training fold, and the observations in the test data were not used for model training or for fitting the calibration function. This procedure prevented information from the test data from entering the probability calibration.

Probabilistic predictive performance was evaluated with the Brier score [7], and calibration, the agreement between predicted probabilities and observed frequencies, was evaluated with reliability analysis. The Brier skill score was also used to characterize the errors of the probabilistic predictions [8, 9]. This allowed us to evaluate discrimination, probabilistic predictive performance and calibration separately under rare-event conditions and to compare the effects of positive-class weighting and post hoc calibration on each.

The predicted probabilities used in the primary analysis and reported in the main text were obtained from the unweighted LightGBM model without class weighting or post hoc isotonic calibration. The positive-class-weighted and isotonically calibrated models were used only in the comparative analysis of rare-event handling and probability calibration.

### 1.3 Operational alert triggering and lead time

To convert predicted probabilities from the road-icing prediction model into operational alerts, the alert condition was defined as met when the predicted probability at each 30-min evaluation time was at or above a set threshold. Two persistence conditions were compared: $k$ = 1, in which the threshold is met at a single evaluation time, and $k$ = 2, in which it is met at both the current and the preceding evaluation time. $k$ = 2 therefore requires the predicted probability to remain at or above the threshold over two evaluation times before an alert is issued; it is not a rule for merging consecutive alerts into a single alert episode.

Operational alert performance was evaluated by setting the probability threshold corresponding to a target POD and then assessing false-alarm frequency and lead time at that detection level. The threshold for each target POD was determined by bisection rather than from a fixed grid of candidate thresholds. An alert was issued when the predicted probability was at or above the threshold, and the highest probability threshold achieving a POD at or above the target was selected as the operational threshold. Predicted probabilities were out-of-fold for the primary observation network and, for the independent observation network, came from models trained on the primary network without retraining; in both cases the threshold itself was chosen on the observed icing outcomes of the evaluation set, so the reported operating points describe the achievable trade-off rather than the performance of a threshold fixed in advance. The same threshold-selection rule was applied to all models, alert settings and observation networks.

False alarms were counted per 30-min evaluation time rather than per alert episode. A false alarm was defined as an evaluation time at which the alert condition was met but no new icing onset occurred within the forecast horizon H, and the count was converted to false alarms per station-day to account for differences in evaluation period among stations. Because there are 48 evaluation times per day, false alarms per station-day were calculated as follows.

$$\mathrm{FA_{day}} = 48 \times \frac{N_{\mathrm{FA}}}{N_{\mathrm{non-event}}} \tag{S1}$$

Here, $N_{\mathrm{FA}}$ is the number of evaluation times at which the alert condition was met without an icing onset within the forecast horizon H, and $N_{\mathrm{non-event}}$ is the total number of evaluation times without an icing onset within that forecast horizon. Consecutive false-alarm evaluation times were counted separately, and no reset or refractory period was applied after icing events. The false-alarm metric in this study is therefore not a count of alert episodes formed by merging persistent alerts but the frequency with which the alert condition is met at non-icing evaluation times, expressed per day.

The lead time for each icing event was defined as the difference between the icing-onset time $t_0$ and the first valid alert time $t_{\mathrm{alert}}$, the first time the alert condition was met for that event.

$$L = t_0 - t_{\mathrm{alert}} \tag{S2}$$

Here, $L > 0$ means that a valid alert was issued before the actual icing onset. POD was the fraction of icing-onset events with at least one valid alert within the preceding forecast horizon, counted over events that had at least one valid evaluation time in that window; event-level lead times were calculated from the first valid alert, so undetected events contribute to POD but not to the lead-time distribution. The forecast horizon H is thus the time window over which the occurrence of future icing onset is predicted at each evaluation time, whereas the lead time L indicates how far the first valid alert preceded the onset of an actual icing event.

This operational evaluation was performed to compare, at the same detection level, the lead time achieved and the associated false-alarm burden, separately from model discrimination as measured by ROC-AUC and PR-AUC.

### 1.4 Cost–loss-based assessment of operational value

The operational value of road-icing predictive information was evaluated with a cost–loss model [10, 11]. With the cost of protective action against icing denoted $C$ and the loss incurred when icing occurs without protective action denoted $L_c$, the cost–loss ratio α was defined as follows.

$$\alpha = \frac{C}{L_c} \tag{S3}$$

In a simple binary decision based on a perfectly calibrated probabilistic forecast, the cost of protective action can be compared with the expected loss without protective action for a predicted icing probability $p$. Under these conditions, taking protective action is optimal when $p \geq \alpha$, and the theoretical decision threshold is $p^* = \alpha$. In this study, however, this threshold was not applied directly as an operational threshold; instead, the potential economic value that the prediction model can provide at different cost–loss ratios was evaluated.

The relative economic value was calculated by normalizing the expected cost of decisions based on predictive information by the expected costs of a reference strategy without predictive information and of a perfect forecast. For a cost–loss ratio $\alpha$, icing frequency s, hit rate H (distinct from the forecast horizon H in Supplementary Methods 1.3) and false alarm rate F, the relative economic value V($\alpha$) was calculated as follows.

$$V(\alpha) = \frac{\min(\alpha,s) - \left[\alpha\left(Hs + F(1-s)\right) + s(1-H)\right]}{\min(\alpha,s) - s\alpha} \tag{S4}$$

Here, $\alpha\left(Hs + F(1-s)\right) + s(1-H)$ is the expected cost of decisions based on predictive information, normalized by the loss $L_c$. The expected cost of the reference strategy without predictive information was defined as $\min(\alpha, s)$, which corresponds to choosing the smaller of the cost $\alpha$ of always taking protective action (always act) and the expected loss $s$ of never taking protective action (never act). The expected cost of a perfect forecast was defined as $s\alpha$ [10]. In this analysis, $s$ was set to 0.0088, the climatological icing frequency for the H = 3 h prediction.

Potential economic value was evaluated over a range of cost–loss ratios and probability thresholds. $\alpha$ was varied from 0.02 to 0.98 in steps of 0.02; for each $\alpha$, the threshold was varied from the 50th to the 99.8th percentile of the predicted probability distribution in steps of 0.2 percentile points, and H and F were calculated for each threshold. For each $\alpha$, the maximum relative economic value was then selected to form the envelope of potential value. The relative economic value reported here is therefore not the realized economic value of a particular pre-specified operational threshold but the maximum potential economic value that the predictive information can provide among the thresholds evaluated at a given cost–loss ratio.

### 1.5 Model interpretation

The direction and relative magnitude of each feature's contribution to predictions of the road-icing prediction model were evaluated with SHAP (SHapley Additive exPlanations) [12]. The SHAP value at each evaluation time was interpreted as a model explanation indicating the direction and magnitude by which the feature raised or lowered the icing prediction relative to the model's baseline prediction. The distributions and magnitudes of SHAP values were compared across features to evaluate how current meteorological and road-surface conditions, recent changes and cumulative antecedent conditions contributed to the predictions of the trained model.

SHAP analysis was used to explain how the trained road-icing prediction model uses each feature. The magnitude and direction of SHAP values therefore represent relative predictive contributions within the model and do not represent causal influences or the magnitude of physical effects of the features on actual road icing.

## Supplementary Methods 2. Construction and selection of features derived from surrounding meteorological fields

### 2.1 Matching objectively analysed fields with RWIS data

To represent meteorological conditions around the RWIS stations spatially, the KMA objectively analysed fields with a spatial resolution of 500 m and a temporal resolution of 5 min were used. The variables were air temperature and dew-point temperature, and data were extracted for the grid cell at each RWIS station and the surrounding grid cells. The objectively analysed fields were matched in time with the RWIS observations to construct the surrounding meteorological fields at icing-onset times and non-icing control times.

## 2.2 Initial spatial features and assessment of spatial scale

The initial analysis used a circular neighbourhood with a radius of 3 km centred on each RWIS station. From the 500 m objective-analysis grid, features representing spatial differences in surrounding air temperature and dew-point temperature were constructed: local anomalies, horizontal gradients, Laplacians and advection-related features. The local anomaly was defined as the difference between the grid cell corresponding to the RWIS station and the mean of the surrounding area, and the horizontal gradient and Laplacian were constructed to represent the spatial change and local curvature of the objectively analysed fields, respectively. Advection-related features combined the spatial gradients of the objectively analysed fields with wind information observed at the RWIS. Temporal changes in air temperature and dew-point temperature were also computed from the current and previous values at the grid cell corresponding to the RWIS station.

To examine whether these fine-scale spatial features provide independent information at the spatial scale of the objectively analysed fields, spatial correlations between grid cells were analysed. Air temperature and dew-point temperature showed very high spatial correlations ($r \geq 0.989$) at all grid separations within 6 km. The final analysis therefore constructed features based on recent temporal changes, surrounding fields over a wider area and upwind conditions instead of horizontal gradients and Laplacians over 500 m–3 km. With the same event–control-time sample and cross-validation splits, the ROC-AUC of the model with the 15 initial features differed from that of the model with the 9 final features by 0.001–0.002.

## 2.3 Final features derived from surrounding meteorological fields

The final features derived from surrounding meteorological fields comprised three categories: temporal changes, surrounding fields and upwind conditions.

First, temporal-change features were constructed from the changes over the last 30 min in air temperature, dew-point temperature and dew-point depression at the objective-analysis grid cell corresponding to each RWIS station. These features captured information on recent cooling or warming and drying or moistening of the air around the road, separately from current meteorological conditions.

Second, surrounding-field features were constructed from objective-analysis grid cells within a 10 km radius of each RWIS station. For air temperature and dew-point temperature, anomaly features were computed as the difference between the value at the grid cell corresponding to the RWIS station and the surrounding 10 km field. These features represented the thermal and moisture state of each RWIS station relative to the surrounding meteorological field.

Third, upwind features were constructed by defining the upwind direction at each time from the wind direction observed at the RWIS, and computing the dew-point depression at grid cells 5, 10 and 20 km upwind and the air temperature difference between the grid cell 10 km upwind and the station grid cell. When the wind direction observation was more than 30 min away from the reference time, the upwind features were set to missing. Wind direction was taken from the nearest 1-min observation at the same station; this was the same minute for 99.998% of evaluation points, and 16 of 1,478,456 points had no observation within 30 min. These features allowed evaluation of the predictive information about future icing onset provided by the moisture conditions of upwind air in addition to conditions at the RWIS station itself.

### 2.4 Feature selection based on the event–control-time analysis

The initial discriminative information of the features derived from surrounding meteorological fields was evaluated with an event–control-time analysis using icing-onset events and non-icing control times. For each of the 2,427 icing-onset events in the primary observation network, one control time was drawn at random from non-icing times at the same station, so that the number of control times equalled the number of events. To reduce the chance that control times included influences before or after icing events, only times at least 12 h away from every icing-onset time were used as control-time candidates. After excluding control times outside the analysis period and times with missing RWIS current-state inputs, the final sample comprised 2,343 icing events and 2,374 control times.

A baseline model using only local current meteorological and road-surface conditions at the RWIS (road surface temperature, air temperature, relative humidity and wind speed) was compared with a model that added the surrounding-field features. LightGBM in this analysis used num_leaves of 15, min_data_in_leaf of 20,

feature_fraction and bagging_fraction of 0.9 and 200 boosting iterations. To reduce overestimation of the additional discriminative information from surrounding-field features caused by overlapping data within particular stations, dates or winters, station-held-out, date-held-out and winter-held-out block cross-validation schemes were applied separately, with each winter defined as November to the following March. These cross-validation schemes tested whether surrounding-field features provided additional discriminative information for distinguishing icing-onset events from non-icing control times under different spatial and temporal data splits.

### 2.5 Application to the road-icing prediction model and past-only analysis

The selected surrounding-field features were joined to the 30-min road-icing prediction evaluation data constructed in the Methods of the main text. The nine features and their spatial scales were fixed in advance from the scale analysis of the full grid and were not reselected within each held-out fold. The road-icing prediction model using only RWIS-based features and the model with surrounding-field features added were compared on the same validation data. Predictive performance was evaluated with ROC-AUC, and operational performance was evaluated using false alarms per station-day and first-trigger lead time at the same target POD. This comparison tested whether surrounding-field features provided additional predictive information for predicting actual future icing onset as well as for event–control-time discrimination.

A past-only analysis was also performed to evaluate whether predictive performance improved because of information shared between same-time objectively analysed fields and RWIS observations. In this analysis, the objectively analysed field at the evaluation time was excluded, and features were defined with reference to a time 1 h or 3 h before the evaluation time. The 30-min change, 10-km-radius anomaly and upwind conditions along the observed wind direction were reconstructed for that time, and the event–control-time evaluation was repeated under station-, date- and winter-held-out cross-validation.

The past-only analysis quantified how much the additional predictive information from surrounding meteorological fields depends on same-time objectively analysed fields. This analysis tested whether additional predictive information was retained when only surrounding meteorological fields available before the evaluation time were used.

## Supplementary Tables

Supplementary Table 1. Critical-slowing-down indicators under two control designs. Six indicators are tested with the same trend statistic (Kendall's $\tau_K$) against random controls and against controls matched on season and hour of day. n = 2,427 icing-onset events and 4,816–4,840 control times. AUC, area under the ROC curve.

| Control design | Indicator | Onset median tauK | Control median tauK | AUC | Fraction tauK > 0 |
|---|---|---|---|---|---|
| Random controls | Rolling variance | -0.103 | -0.077 | 0.501 | 0.461 |
| Random controls | Lag-1 autocorrelation | 0.026 | -0.051 | 0.526 | 0.509 |
| Random controls | Skewness | -0.128 | 0.000 | 0.431 | 0.401 |
| Random controls | Kurtosis | 0.026 | -0.026 | 0.532 | 0.508 |
| Random controls | Spectral reddening | 0.051 | -0.077 | 0.538 | 0.519 |
| Random controls | Autocorrelation decay time | -0.092 | -0.051 | 0.481 | 0.443 |
| Season- and hour-matched | Rolling variance | -0.103 | -0.256 | 0.548 | 0.461 |
| Season- and hour-matched | Lag-1 autocorrelation | 0.026 | -0.154 | 0.560 | 0.509 |
| Season- and hour-matched | Skewness | -0.128 | -0.051 | 0.455 | 0.401 |
| Season- and hour-matched | Kurtosis | 0.026 | -0.026 | 0.534 | 0.508 |
| Season- and hour-matched | Spectral reddening | 0.051 | -0.154 | 0.567 | 0.519 |
| Season- and hour-matched | Autocorrelation decay time | -0.092 | -0.105 | 0.510 | 0.443 |

Supplementary Table 2. Positive control using synthetic series with declining recovery rates. The same analysis procedure is applied to a linear stochastic process whose recovery rate declines towards the end of the window. The procedure begins to separate synthetic transitions from synthetic controls once the lag-1 autocorrelation rises by about 0.02 (AUC 0.549 and 0.555); the discrimination observed for icing onsets (0.488 and 0.513) is at the level of settings with a rise of 0.01 or less (0.508 and 0.515).

| Setting | lambda at window end | AR1 at window end | AR1 rise | AUC (variance trend) | AUC (AR1 trend) | Fraction tauK > 0 | Windows |
|---|---|---|---|---|---|---|---|
| No missing data | 0.005 | 0.995 | 0.045 | 0.618 | 0.636 | 0.652 | 600 |
| No missing data | 0.010 | 0.990 | 0.040 | 0.600 | 0.608 | 0.627 | 600 |
| No missing data | 0.020 | 0.980 | 0.030 | 0.584 | 0.585 | 0.638 | 600 |
| No missing data | 0.030 | 0.970 | 0.020 | 0.549 | 0.555 | 0.578 | 600 |
| No missing data | 0.040 | 0.960 | 0.010 | 0.508 | 0.515 | 0.510 | 600 |
| No missing data | 0.050 | 0.950 | 0.000 | 0.512 | 0.507 | 0.500 | 600 |
| Block missing 5% | 0.005 | 0.995 | 0.045 | 0.631 | 0.635 | 0.668 | 600 |
| Block missing 10% | 0.005 | 0.995 | 0.045 | 0.618 | 0.629 | 0.643 | 600 |
| Block missing 10% | 0.020 | 0.980 | 0.030 | 0.561 | 0.563 | 0.573 | 600 |

Supplementary Table 3. Alternative state variables. The same test is applied to four variables besides road surface temperature. Critical slowing down predicts that variance and autocorrelation rise together; for the two variables that measure the surface state directly, the two indicators move in opposite directions.

| **State variable** | **Indicator** | **AUC** | **Fraction tauK > 0** | **Onset windows** | **Control windows** |
|---|---|---|---|---|---|
| Surface temperature Ts (baseline) | Rolling variance | 0.488 | 0.461 | 2427 | 4837 |
| Surface temperature Ts (baseline) | Lag-1 autocorrelation | 0.513 | 0.509 | 2427 | 4837 |
| Deposition margin Mdep = Ts − Tf | Rolling variance | 0.502 | 0.466 | 2341 | 4734 |
| Deposition margin Mdep = Ts − Tf | Lag-1 autocorrelation | 0.567 | 0.554 | 2341 | 4734 |
| Air temperature Ta | Rolling variance | 0.471 | 0.460 | 2341 | 4734 |
| Air temperature Ta | Lag-1 autocorrelation | 0.549 | 0.564 | 2341 | 4734 |
| Friction coefficient FRC | Rolling variance | 0.683 | 0.668 | 2425 | 4829 |
| Friction coefficient FRC | Lag-1 autocorrelation | 0.308 | 0.315 | 2425 | 4829 |
| Water-film thickness WFT | Rolling variance | 0.757 | 0.854 | 595 | 1465 |

| Water-film thickness WFT | Lag-1 autocorrelation | 0.231 | 0.183 | 595 | 1465 |
|---|---|---|---|---|---|

Supplementary Table 4. Lead-time dependence of the indicator trends. The end of the analysis window is moved back from icing onset by the stated interval. Discrimination by the friction-coefficient variance trend decreases as the analysis window ends earlier, which is more consistent with a change that accompanies the transition than with one that precedes it.

| **Variable** | **Indicator** | **0 min** | **30 min** | **60 min** | **120 min** | **180 min** | **360 min** | **720 min** |
|---|---|---|---|---|---|---|---|---|
| Friction coefficient FRC | Rolling variance | 0.679 | 0.650 | 0.636 | 0.614 | 0.601 | 0.563 | 0.543 |
| Friction coefficient FRC | Lag-1 autocorrelation | 0.311 | 0.349 | 0.370 | 0.395 | 0.407 | 0.442 | 0.460 |
| Surface temperature Ts | Rolling variance | 0.501 | 0.489 | 0.481 | 0.473 | 0.469 | 0.475 | 0.558 |
| Surface temperature Ts | Lag-1 autocorrelation | 0.528 | 0.515 | 0.507 | 0.494 | 0.486 | 0.482 | 0.520 |

Supplementary Table 5. Sensitivity of CSD results to inclusion of the incomplete 2022/23 winter. The 2022/23 winter was incomplete because observations began on 29 December 2022 and only 2–6 RWIS stations were operational during this period. Including this winter adds two icing-onset events and changes the discrimination AUC by at most 0.004, which is smaller than the spread obtained by redrawing the non-icing control times alone (up to 0.009); the choice of analysis period therefore has negligible influence on the CSD results.

| **Analysis period** | **Icing-onset events** | **Variance-trend AUC** | **AR1-trend AUC** |
|:---:|---:|---:|---:|
| Primary period (2023/24–2025/26) | 2,427 | 0.488 | 0.513 |
| Including 2022/23 winter | 2,429 | **0.489** | **0.517** |
| Absolute change | +2 events | +0.0005 | +0.0036 |

Supplementary Table 6. External application to the independent station set. The road-icing prediction model trained on the primary observation network is applied unchanged to the 127 stations that share no station with it. Discrimination is lower

than in cross-validation on the primary network but remains well above the no-discrimination value of 0.5.

| Forecast lead | Set | AUC | Base rate | POD | False alarms per station-day | First-trigger lead (median) |
|---|---|---|---|---|---|---|
| T+3 h | Independent 127 stations (external application) | 0.849 | 0.0138 | 0.80 | 6.59 | 153 min |
| T+6 h | Independent 127 stations (external application) | 0.820 | 0.0243 | 0.80 | 6.32 | 266 min |

Supplementary Table 7. Operational alert performance for every configuration reported in the main text, including the comparison with and without surrounding-field features that Fig. 6b summarises. All operating points are determined by a single rule, the largest threshold satisfying POD ≥ target found by bisection, so the reported false alarms per station-day do not depend on a grid of candidate thresholds. POD, probability of detection; IQR, interquartile range.

| Configuration | Forecast lead | Persistence k | Target POD | Achieved POD | False alarms per station-day | First-trigger lead (median) | IQR |
|---|---|---|---|---|---|---|---|
| T+6, POD 0.80 | T+6 h | 1 | 0.80 | 0.80 | 3.57 | 246 min | 124–336 min |
| T+6, POD 0.70 | T+6 h | 1 | 0.70 | 0.70 | 2.14 | 229 min | 108–332 min |
| T+3, POD 0.80 | T+3 h | 1 | 0.80 | 0.80 | 3.96 | 151 min | 85–165 min |
| T+6, persistence k = 2, POD 0.55 | T+6 h | 2 | 0.55 | 0.55 | 1.47 | 208 min | 89–319 min |
| T+6, RWIS only | T+6 h | 1 | 0.80 | 0.80 | 3.57 | 246 min | 124–336 min |
| T+6, RWIS + surrounding fields | T+6 h | 1 | 0.80 | 0.80 | 1.17 | 203 min | 99–305 min |
| T+3, RWIS only | T+3 h | 1 | 0.80 | 0.80 | 3.96 | 151 min | 85–165 min |
| T+3, RWIS + surrounding fields | T+3 h | 1 | 0.80 | 0.80 | 1.50 | 128 min | 69–161 min |

| Model 0 (temperature threshold) | T+3 h | 1 | 0.80 | 0.803 | 15.69 | 157 min | 113–168 min |
|---|---|---|---|---|---|---|---|
| Model 1 (current state) | T+3 h | 1 | 0.80 | 0.80 | 6.72 | 147 min | 73–164 min |
| Model 2 (current + recent + cumulative) | T+3 h | 1 | 0.80 | 0.80 | 3.96 | 151 min | 85–165 min |

**Supplementary References**